\documentclass[aps,prd,11pt,superscriptaddress,notitlepage,longbibliography,nofootinbib,tightenlines]{revtex4-1}

\usepackage{graphicx,amsmath,amssymb,multirow,array,bm,mathrsfs}
\usepackage{epsf,amsfonts,bbold,color}

\usepackage[pdftex]{hyperref}
\hypersetup{colorlinks=true, linkcolor=darkred, citecolor=blue, linktoc=page}
\definecolor{darkred}{rgb}{0.8,0.1,0.1}

\usepackage{slashed}
\usepackage{braket}

\newcommand{\N}{\mathcal{N}}

\newcommand{\beq}{\begin{equation}}
\newcommand{\eeq}{\end{equation}}

\newcommand{\tr}{{\rm{tr}}}

\newcommand{\bea}{\begin{eqnarray}}
\newcommand{\eea}{\end{eqnarray}}
\newcommand{\bdm}{\begin{displaymath}}
\newcommand{\edm}{\end{displaymath}}
\newcommand{\bal}{\begin{align}}
\newcommand{\eal}{\end{align}}
\newcommand{\nn}{\nonumber}

\newcommand{\vol}{\text{vol}}

\newcommand{\p}{\pi}
\newcommand{\tentsp}{T_{\textrm{ent}}^{\textrm{sphere}}}
\newcommand{\tentst}{T_{\textrm{ent}}^{\textrm{strip}}}

\def\le{\left}
\def\ri{\right}

\def\pa{\partial}

\def\le{\left}
\def\ri{\right}
\def\tr{{\rm Tr}}

\def\a{\alpha}
\def\b{\beta}
\def\g{\gamma}
\def\d{\delta}

\def\l{\lambda}

\def\f{\phi}
\def\e{\epsilon}

\def\m{\mu}
\def\n{\nu}

\def\y{\psi}

\def\ca{\mathcal{A}}

\def\co{\mathcal{O}}

\def\G{\Gamma}

\include{macros}

\def\tent{T_\mathrm{ent}}

\def\cw{\mathcal{W}}
\def\ce{\mathcal{E}}
\def\see{S_\mathrm{EE}}

\newcommand{\T}[1]{\langle T_{#1} \rangle}
\newcommand{\gsup}[2]{g^{(#1)}_{#2}}
\newcommand{\hsup}[2]{h^{(#1)}_{#2}}

\begin{document}

\title{A First Law of Entanglement Rates from Holography}

\author{Andy O'Bannon}
\email{a.obannon@soton.ac.uk}
\affiliation{STAG Research Centre, Physics and Astronomy, University of Southampton, \\ \hspace{.3cm} Highfield, Southampton SO17 1BJ, U.~K.}

\author{Jonas Probst}
\email{Jonas.Probst@physics.ox.ac.uk}
\affiliation{Rudolf Peierls Centre for Theoretical Physics, University of Oxford, \\ \hspace{.3cm} 1 Keble Road, Oxford OX1 3NP, U.~K.}

\author{Ronnie Rodgers}
\email{R.J.Rodgers@soton.ac.uk}
\affiliation{STAG Research Centre, Physics and Astronomy, University of Southampton, \\ \hspace{.3cm} Highfield, Southampton SO17 1BJ, U.~K.}

\author{Christoph F.~Uhlemann}
\email{uhlemann@physics.ucla.edu}

\affiliation{Department of Physics, University of Washington, Seattle, WA 98195-1560, USA}
\affiliation{Mani L. Bhaumik Institute for Theoretical Physics, Department of Physics and Astronomy, \\ University of California, Los Angeles, CA 90095, USA}

\begin{abstract}
For a perturbation of the state of a Conformal Field Theory (CFT), the response of the entanglement entropy is governed by the so-called ``first law'' of entanglement entropy, in which the change in entanglement entropy is proportional to the change in energy. Whether such a first law holds for other types of perturbations, such as a change to the CFT Lagrangian, remains an open question. We use holography to study the evolution in time $t$ of entanglement entropy for a CFT driven by a $t$-linear source for a conserved $U(1)$ current or marginal scalar operator. We find that although the usual first law of entanglement entropy may be violated, a first law for the rates of change of entanglement entropy and energy still holds. More generally, we prove that this first law for rates holds in holography for any asymptotically $(d+1)$-dimensional Anti-de Sitter metric perturbation whose $t$ dependence first appears at order $z^d$ in the Fefferman-Graham expansion about the boundary at $z=0$. 
\end{abstract}

\maketitle

\tableofcontents

\section{Introduction, Summary, and Outlook}
\label{intro}

Many-body systems in thermal equilibrium are governed by universal laws, the laws of thermodynamics. Many-body systems perturbed out of thermal equilibrium are also governed by universal laws, the laws of hydrodynamics (for sufficiently small perturbations and at sufficiently late times). What laws, if any, govern many-body systems driven far from equilibrium? This question is of central importance in many branches of physics, from cosmology (the electroweak phase transition, the Kibble-Zurek mechanism, etc.) to condensed matter physics (quantum quenches, thermalization, etc.) to heavy ion collisions (thermalization and isotropization of the quark-gluon plasma), and beyond. Many of these phenomena, such as thermalization, necessarily involve interactions. Few reliable techniques exist for studying interacting systems far from equilibrium, hence the question remains open.

Cardy and Calabrese pioneered the use of entanglement entropy (EE), $\see$, to characterize far-from-equilibrium systems~\cite{Calabrese:2005in,Calabrese:2007mtj,Calabrese:2009qy,Calabrese:2016xau}. The EE of a sub-region of space at a fixed time, $t$, is defined as the von Neumann entropy of the reduced density matrix, $\rho$, obtained by tracing out the states in the rest of space (\textit{i.e.}\ the region's complement), $\see \equiv - \textrm{tr} \left(\rho \ln \rho\right)$.

Cardy and Calabrese focused on a quantum quench of a coupling in the Hamiltonian to a value that produces a Conformal Field Theory (CFT), and used the powerful techniques of (boundary) CFT in spacetime dimension $d=2$ to compute $\see$, for a spatial interval of length $\ell$. They showed that after the quench ends, $\see$ evolves linearly in $t$, and then saturates at a time proportional to $\ell/c$, with $c$ the speed of light. They also provided an intuitive model for $\see$'s evolution, in terms of maximally-entangled (EPR) pairs of particles produced by the quench, which are necessarily massless, due to the CFT's scale invariance, and hence move at speed $c$. Liu and Suh proposed, based on evidence from the Anti-de Sitter/CFT (AdS/CFT) correspondence, also known as holography, that when $d>2$, Cardy and Calabrese's massless particle model becomes an ``entanglement tsunami'' in which a quench produces a wave-front of entangled excitations that moves inward from the region's boundary~\cite{Liu:2013iza,Liu:2013qca,Casini:2015zua}.

Crucially, EE obeys constraints that ultimately come from unitarity, and that can be, and have been, used to constrain far-from-equilibrium evolution in quantum systems. For example, for two density matrices $\rho$ and $\rho'$ in the same Hilbert space, their relative entropy, $S(\rho|\rho')\equiv \textrm{tr}\left(\rho\ln \rho\right)-\textrm{tr}\left(\rho\ln \rho'\right)$, is non-negative, and indeed provides a measure of the ``statistical distance'' or distinguishability between them. Positivity of relative entropy, $S(\rho|\rho')\geq 0$, played a key role in proving speed limits on entanglement tsunamis~\cite{Liu:2013iza,Liu:2013qca,Avery:2014dba,Casini:2015zua,Hartman:2015apr}.

Further constraints can be derived from $S(\rho|\rho')\geq0$. In particular, if $\rho$ and $\rho'$ are close, so that $\rho' = \rho + \delta \rho$ with $\delta \rho$ small, then expanding $S(\rho|\rho')$ to first order in $\delta \rho$ gives a constraint called the ``First Law of EE" (FLEE)~\cite{Bhattacharya:2012mi,Allahbakhshi:2013rda,Blanco:2013joa,Wong:2013gua}:
\beq
\label{flee1}
\delta \see = \delta \langle H\rangle,
\eeq
where $\delta \see$ is the change in EE, $H$ is the modular Hamiltonian, defined via $\rho \equiv e^{-H}$, and $\delta \langle H \rangle$ is the change in $H$'s expectation value. If $\rho$ is a thermal density matrix with temperature $T$, then the FLEE becomes the usual first law of thermodynamics, $\delta S = \delta E/T$, with entropy $S$ and energy $E$. If $\rho$ is the reduced density matrix of a spatial sub-region, then generically $H$ and $\delta \langle H\rangle$ are complicated non-local objects that are difficult to calculate. However, for a spherical sub-region in a CFT vacuum, $H$ is a product of two factors, integrated over the sphere's volume. The first factor is $\delta E$, or in terms of the stress-energy tensor $T_{\mu\nu}$ ($\mu,\nu=0,1,\ldots,d-1$, with $x^0\equiv t$), the change in $\langle T_{tt}\rangle$ in the sub-region. The second factor depends only on geometric data, including in particular the sphere's radius, $R$~\cite{Casini:2011kv,Wong:2013gua}. For perturbations with $\delta \langle T_{tt}\rangle$ constant in space, or for spheres sufficiently small that $\delta\langle T_{tt}\rangle$ can be approximated as constant, the integral is easily performed, with the result
\beq
\label{flee2}
\delta \see = \frac{\delta E}{\tent},
\eeq
where the ``entanglement temperature,'' $\tent$, depends on $R$ and $d$,
\beq
\label{spheretent}
\tentsp = \frac{d+1}{2 \pi R},
\eeq
but is independent of any other details of the CFT or of the state $\rho$. For a ``strip'' sub-region, consisting of two parallel planes separated by a distance $\ell$, holographic CFTs also obey eq.~\eqref{flee2}, but now with~\cite{Bhattacharya:2012mi}
\beq
\label{striptent}
\tentst 	= \frac{2\,(d^2-1) \,\Gamma\left(\frac{d+1}{2(d-1)}\right) \Gamma\left(\frac{d}{2(d-1)}\right)^2}{\sqrt{\pi}\,\Gamma\left(\frac{1}{d-1}\right) \Gamma\left(\frac{1}{2(d-1)}\right)^2} \,\frac{1}{\ell}.
\eeq

The FLEE does not hold for arbitrary deformations. In quantum mechanics, the FLEE holds only for ``completely positive trace-preserving'' maps, linear maps that are combinations of unitary transformations, partial tracing, and adding sub-systems---for a precise definition, see for example appendix A of ref.~\cite{Blanco:2013joa}, and references therein.

In a continuum Quantum Field Theory (QFT), what deformations obey the FLEE? Finding a precise answer appears to be more challenging than in quantum mechanics. In particular, in continuum QFT, $\rho$ generically has an infinite number of eigenvalues, so in what sense can a perturbation of the eigenvalues, $\delta \rho$, be small? Currently the best intuition appears to be that, for compact sub-regions, the FLEE should hold when $\delta\langle T_{\mu\nu}\rangle$ is small, relative to the scale set by the sub-region's size~\cite{Blanco:2013joa}.  

In this paper we will consider perturbations that go beyond a change of state: we will deform a CFT Hamiltonian by a relevant or marginal operator with a $t$-dependent source, which drives the CFT far from equilibrium. We will focus on sources linear in $t$, although our most general results apply to a larger class of sources, characterized most precisely via holography, as we discuss below. For our cases, we will show two things: first, generically the na\"ive FLEE in eq.~\eqref{flee2} is violated, and second, a relation very similar to eq.~\eqref{flee2} holds for the \textit{rates of change} of EE and energy.

We will restrict to CFTs with holographic duals, mainly because holography is currently the easiest way to compute $\see$ in interacting QFTs. Computing $\see$ holographically requires two steps. First, we must solve Einstein's equation for the asymptotically-AdS$_{d+1}$ metric, $G_{mn}$ ($m,n = 0,1,\ldots,d$), of the holographically dual spacetime. We will mostly work with a Fefferman-Graham (FG) holographic coordinate $z$ with asymptotic $AdS_{d+1}$ boundary at $z=0$, where the CFT ``lives.'' Second, we must compute the area of the extremal surface that at the asymptotic $AdS_{d+1}$ boundary coincides with the entangling region's boundary in the dual QFT. $\see$ is then that area divided by $4G_N$, with Newton's constant $G_N$~\cite{Ryu:2006bv,Ryu:2006ef,Hubeny:2007xt,Lewkowycz:2013nqa,Dong:2016hjy}.

In holography, a deformation of the CFT Hamiltonian by a relevant or marginal operator corresponds to a change of the bulk metric, $G_{mn} \to G_{mn} + \delta G_{mn}$. Our unperturbed metric $G_{mn}$ will be asymptotically $AdS_{d+1}$ and independent of $t$ and the CFT spatial coordinates, but otherwise arbitrary. Our main examples of $G_{mn}$ will be Poincar\'e patch $AdS_{d+1}$, dual to the CFT vacuum, and the $AdS_{d+1}$ black brane, dual to the CFT with non-zero $T$. Our perturbation $\delta G_{mn}$ will preserve the asymptotic $AdS_{d+1}$, but generically depend on $t$. Our only non-trivial assumption will be that $t$-dependence in $\delta G_{mn}$ first appears at order $z^d$ in the FG expansion. In other words, the $t$ dependence of $\delta G_{mn}$ will be arbitrary, except terms in the FG expansion with powers of $z$ smaller than $z^d$ will be $t$-independent. With that assumption, in section~\ref{general} we will prove a ``First Law Of Entanglement Rates'' (FLOER),
\beq
\label{floer}
\partial_t \delta \see = \frac{\partial_t \delta E}{\tent},
\eeq
where $\tent$ depends on the unperturbed $G_{mn}$ and the extremal surface therein. If the unperturbed $G_{mn}$ is Poincar\'e patch $AdS_{d+1}$, then $\tent$ is identical to that in eq.~\eqref{spheretent} or~\eqref{striptent}.

Our proof of eq.~\eqref{floer} can also be straightforwardly extended to deformations by sources which are position-dependent instead of time-dependent, provided the corresponding assumptions about $G_{mn}$ and $\delta G_{mn}$ are satisfied. The resulting FLOER involves rates of change in a spatial coordinate $x^1\equiv x$, rather than $t$, \textit{i.e.}\ $\partial_x \delta \see = \partial_x \delta E/\tent$. However, given our motivation to understand far-from-equilibrium evolution, and also for clarity, we will continue to refer only to $t$-dependent sources, unless stated otherwise.

Eq.~\eqref{floer} is our main result. The key assumption underlying eq.~\eqref{floer}, that $t$-dependence in $\delta G_{mn}$ appears first at order $z^d$ in the FG expansion, characterizes the most general class of perturbations for which our FLOER holds, and turns out to be a relatively mild constraint. Indeed, in sections~\ref{vaidya},~\ref{d3d7}, and~\ref{massless} we discuss various non-trivial examples that illustrate how easily our key assumption can be satisfied with a $t$-linear source. Our examples also provide our other main result: in many of our examples the FLEE in eq.~\eqref{flee2} is explicitly violated, indicating that the FLOER may be more fundamental than the FLEE, as we discuss below.

In sections~\ref{vaidya} and~\ref{d3d7}, we consider holographic CFTs in $d=3$ and $4$, respectively, each with a conserved current $J^{\mu}$ of a global $U(1)$ symmetry. In each case, in the CFT we introduce a constant external electric field $\cal{E}$ in the $x$ direction, that is, we add to the CFT Lagrangian a relevant deformation $\propto t\, \mathcal{E}J^x$, resulting in a current, $\langle J^x \rangle \neq 0$. We introduce no charge density, $\langle J^t \rangle =0$, so the current arises exclusively from Schwinger pair production, \textit{i.e.} production of maximally-entangled particle-anti-particle (EPR) pairs. Crucially, in our examples, $\langle J^x \rangle$ is $t$-independent. As a result, the Ward identity $\partial_{\mu} T^{\mu\nu} = F^{\nu\rho} J_{\rho}$ implies Joule heating, $\partial_t \langle T_{tt}\rangle = \mathcal{E} \langle J^x \rangle$, that is also $t$-independent. As a convenient shorthand, we will call such states ``Non-Equilibrium Steady-States'' (NESS): non-equilibrium because $\partial_t \langle T_{tt}\rangle\neq 0$, but steady states because $\langle J^x \rangle$ and $\partial_t \langle T_{tt}\rangle$ are $t$-independent.

In holography, $T_{\mu\nu}$ is dual to $G_{mn}$, and $J^{\mu}$ is dual to a $U(1)$ gauge field, $A_m$. On the gravity side of the duality, our examples in sections~\ref{vaidya} and~\ref{d3d7} thus both have $G_{mn}$ and $A_m$, albeit with some essential differences.

In section~\ref{vaidya}, we consider Einstein-Maxwell theory in $AdS_4$, which arises for example from the consistent truncation of eleven-dimensional supergravity on $AdS_4 \times S^7$ down to $AdS_4$~\cite{Chamblin:1999tk,Cvetic:1999xp}. In that example, the dual CFT is the ABJM theory, \textit{i.e.}\ the $\N=6$ supersymmetric (SUSY) Chern-Simons-matter CFT in $d=3$~\cite{Aharony:2008ug}. Our NESS are dual to spacetimes with a null $U(1)$ field strength and $AdS_4$-Vaidya metric~\cite{Horowitz:2013mia}, describing a horizon that moves towards the asymptotically $AdS_4$ boundary as $t$ increases.

In contrast, in section~\ref{d3d7} our $A_m$ has a probe Dirac-Born-Infeld (DBI) action in a fixed asymptotically $AdS_5$ background. Specifically, we consider asymptotically $AdS_5 \times S^5$ solutions of type IIB supergravity with a number $N_f$ of probe D7-branes along $AdS_5 \times S^3$. The type IIB solutions are dual to states of $\N=4$ $SU(N_c)$ SUSY Yang-Mills (SYM) theory in $d=4$, at large $N_c$ and large 't Hooft coupling, and the probe D7-branes are dual to a number $N_f\ll N_c$ of $\N=2$ SUSY hypermultiplets in the fundamental representation of $SU(N_c)$, \textit{i.e.}\ flavor fields. When $T \neq 0$, $\langle T^{\mu\nu}\rangle$ receives order $N_c^2$ and $N_f N_c$ contributions from the $\N=4$ SYM and flavor fields, respectively. We may thus think of the flavors as probes inside a huge heat bath. Our NESS exist because $\mathcal{E}$ pumps energy into the flavor sector at the same constant rate that the flavors dissipate energy into the heat bath. To obtain $\delta \see$ we compute only the linearized (not the full non-linear) back-reaction of $A_m$ onto $G_{mn}$.

Although in sections~\ref{vaidya} and~\ref{d3d7} we focus on particular ``top-down'' string/M-theory constructions, in each case our analysis should easily generalize to many other systems of $U(1)$ gauge fields in asymptotically $AdS_{d+1}$ spacetimes, either fully back-reacted, as in section~\ref{vaidya}, or with linear back-reaction of a probe, as in section~\ref{d3d7}.

In section~\ref{massless}, we consider holographic CFTs in $d=2,3,4$ with a marginal scalar operator $\mathcal{O}$, and add to the CFT Lagrangian a deformation $\propto t \, \mathcal{O}$. In holography, a marginal $\mathcal{O}$ is dual to a massless scalar field, $\phi$. In section~\ref{massless} we compute only $\phi$'s linearized back-reaction onto $G_{mn}$, and only in the asymptotically $AdS_{d+1}$ region, which suffices to establish the FLOER. (The appendix contains the results of the holographic renormalisation~\cite{deHaro:2000vlm} of $\phi$ in $d=3$ and $4$ that we use in section~\ref{massless}.) In section~\ref{massless} we also follow ref.~\cite{Andrade:2013gsa}, and add to the CFT Lagrangian a deformation $\propto x \, \mathcal{O}$. In that case, a spatial FLOER is satisfied trivially, because in the system of ref.~\cite{Andrade:2013gsa} both $\delta \see$ and $\delta E$ turn out to be $x$-independent.

In our examples symmetries actually require $\mathcal{T}_{mn}$ to depend only on $z$, and not on $t$. In sections~\ref{vaidya} and~\ref{d3d7}, $U(1)$ gauge invariance implies that $\mathcal{T}_{mn}$ depends only on $A_m$'s field strength, $F_{mn}$, which is $t$-independent because our solutions for $A_m$ are linear in $t$. In section~\ref{massless}, the massless scalar $\phi$ has a shift symmetry $\phi \to \phi + C$ with constant $C$, which implies $\mathcal{T}_{mn}$ depends only on derivatives of $\phi$, and hence is $t$-independent because our solutions for $\phi$ are linear in $t$. Time-dependence is instead generated by off-diagonal terms $\mathcal{T}_{tz}=\mathcal{T}_{zt}$ which, via Einstein's equation, force $\delta G_{mn}$ to depend on both $z$ and $t$. Indeed, such off-diagonal terms in $\mathcal{T}_{mn}$ indicate $\partial_t \langle T_{tt}\rangle \neq 0$ in the dual QFT~\cite{Karch:2008uy}, \textit{i.e.}\ the system is out of equilibrium. We emphasize, however, that while the symmetries of our examples are \textit{sufficient} to guarantee that $\delta G_{mn}$ obeys our key assumption, they are not strictly \textit{necessary}.

In terms of the CFT generating functional, in all of our examples we deform the CFT by a source \textit{linear} in $t$. Such deformations are \textit{not} quenches in any conventional sense: our systems do not necessarily approach equilibrium in the infinite past or future. At best, our deformations could perhaps be interpreted as an endless series of global quenches, one right after another, every moment in $t$. More succinctly, our systems are \textit{driven} by a source linear in $t$ (not periodic in $t$, in contrast to ref.~\cite{Rangamani:2015sha}). We emphasize again, however, that our examples are only a subset of a much larger class of $t$-dependent deformations, as mentioned above.

To summarize, we have identified a law governing a certain class of far-from-equilibrium systems. Specifically, we extended the FLEE in eq.~\eqref{flee2} beyond deformations of the state, to deformations of the Hamiltonian, characterized holographically by $\delta G_{mn}$ whose $t$-dependence first appears at order $z^d$ in the FG expansion. For such deformations, we have shown that the FLOER of eq.~\eqref{floer} holds, while the FLEE of the form in eq.~\eqref{flee2} in general does not.

Looking to the future, our results have implications both practical and conceptual. In practical terms, the FLOER may be useful because $\partial_t \delta E$ is often easier to calculate than $\partial_t\delta \see$. In particular, if we can argue that the FLOER holds, and we know $\tent$, then we can obtain $\partial_t \delta \see$ by calculating $\partial_t \delta E$, for example via the Ward identity $\partial_{\mu} T^{\mu\nu} = F^{\nu\rho} J_{\rho}$.

Of the many conceptual questions our results raise, we will highlight only three. First, given that the same $\tent$ appears in our FLOER and in the FLEE of eq.~\eqref{flee2}, can the FLOER simply be integrated to obtain the FLEE? In our examples where the FLEE is violated, $\delta \see$ has a $t$-independent contribution absent from $\delta E$. Apparently, integrating the FLOER produces different integration constants in $\delta \see$ and $\delta E$. We suspect that the difference arises from initial conditions. For instance, imagine ``turning on'' our $t$-linear source at $t=0$. We expect EE and energy to be produced immediately. However, the EE is only sensitive to entanglement across the entangling surface, so in an entanglement tsunami description some of the EPR pairs produced at $t=0$ will contribute to $\see$ only after some ``lag time'' required for one EPR partner to leave the sub-region. The lag time should be on the order of the sub-region's size, as indeed we find in some of our examples.

Second, when the FLOER holds but the FLEE in the form of eq.~\eqref{flee2} is violated, could the FLEE in the form of eq.~\eqref{flee1} still hold? This is only possible if $\delta \langle H\rangle \neq \delta E/\tent$. The crucial point is that we are not comparing two states in the same Hilbert space. We are changing the CFT Hamiltonian, which changes the Hilbert space, and then comparing states in the old and new Hilbert spaces. In such cases, can $S(\rho|\rho')$ even be \textit{defined}, and if so, do $S(\rho|\rho')\geq0$ and hence the FLEE in eq.~\eqref{flee1} hold? To our knowledge, these questions remain open. The current state of the art appears to be the proof in ref.~\cite{Casini:2016udt}, for $t$-\textit{independent} relevant deformations, that $S(\rho|\rho')$ can be defined, and $S(\rho|\rho')\geq 0$, for states in two different Hilbert spaces only if the two theories have the same UV fixed point\footnote{See also ref.~\cite{Carracedo:2016qrf} for a discussion of whether $S(\rho|\rho')\geq 0$ holds for a deformation $\propto t \, \mathcal{O}$, with marginal $\mathcal{O}$, for CFTs on a spatial sphere, holographically dual to gravity in \textit{global} $AdS_{d+1}$.}. The D3/D7 system with massive flavors actually provides a \textit{time-independent} example where the assumptions of ref.~\cite{Casini:2016udt} are satisfied but the FLEE in the form of eq.~\eqref{flee2} fails, as we discuss in sec.~\ref{d3d7}. In our \textit{time-dependent} examples we could attempt to test the FLEE in eq.~\eqref{flee1} directly, by calculating $\delta \langle H\rangle$ holographically. However, although much is known about the holographic dual of $H$~\cite{Czech:2012bh,Headrick:2014cta,Jafferis:2014lza,Jafferis:2015del,Dong:2016eik}, we know of no practical prescription for computing $\delta \langle H \rangle$ holographically, so we will leave such a test for future research.

Third, can we identify more precisely in field theory terms the class of $t$-dependent deformations for which the FLOER of eq.~\eqref{floer} holds while the FLEE of eq.~\eqref{flee2} need not? Moreover, can we extend our results to more general systems, either in QFT or in holography (for work in this direction, see for example ref.~\cite{Lokhande:2017jik})? We believe that these and many other questions relating to the FLOER deserve further study, in large part because they may eventually reveal universal laws governing far-from-equilibrium systems.

\section{General Analysis}
\label{general}

In this paper we consider only asymptotically $AdS_{d+1}$ spacetimes. In this section, we exclusively use FG coordinates, in which the metric takes the form
\beq
\label{eq:fefferman_graham_form}
d s^2 = G_{mn}dx^m dx^n = \frac{L^2}{z^2}(dz^2 + g_{\m\n}(z,x^{\rho}) dx^\m dx^\n),
\eeq
where $m,n = 0,1,\ldots,d$ and $\mu,\nu,\rho = 0,1,\ldots,d-1$, where $x^0=t$ is time, and $L$ is the radius of the asymptotic $AdS_{d+1}$, with boundary at $z = 0$. The FG expansion of $g_{\mu\nu}(z,x^{\rho})$ about the boundary is of the form
\beq
\label{eq:metric_expansion}
g_{\m\n}(z,x^{\rho}) = \gsup{0}{\mu\nu}(x^{\rho}) + z^2 \, \gsup{2}{\mu\nu}(x^{\rho}) + \ldots + z^d \, \gsup{d}{\m\n}(x^{\rho}) + z^{d} \log z^2\,\hsup{d}{\m\n}(x^{\rho}) + \ldots,
\eeq
where the term $\propto z^{d} \log z^2$ is present only when $d$ is even. The expectation value of the energy-momentum (density) tensor of the dual field theory, $\langle T_{\m\n}(x^{\rho})\rangle$, takes the generic form~\cite{deHaro:2000vlm}
\beq
\label{eq:general_energy_momentum}
\langle T_{\m\n}(x^{\rho})\rangle = \frac{d L^{d-1}}{16\p G_N}\,\gsup{d}{\m\n}(x^{\rho}) + X_{\m\n}[\gsup{N}{\kappa\lambda}(x^{\rho})],
\eeq
where $X_{\m\n}[\gsup{N}{\kappa\lambda}(x^{\rho})]$ is a function of the $\gsup{N}{\kappa\lambda}(x^{\rho})$ with $N<d$. Via Einstein's equation, the $\gsup{N}{\m\n}(x^{\rho})$ with $N<d$ are functions of the leading asymptotic coefficients in the near-boundary FG expansions of matter fields, or in dual QFT terms, functions of sources of operators.

Our key assumption is that the $\gsup{N}{\m\n}(x^{\rho})$ with $N<d$ are $t$-independent: $\gsup{N}{\m\n}(x^{\rho}) = \gsup{N}{\m\n}(\vec{x})$, where $\vec{x}$ are the field theory spatial coordinates. In these cases,
\beq
\label{eq:energy_momentum_derivative}
\partial_t \langle T_{\m\n}(x^{\rho})\rangle = \frac{d L^{d-1}}{16\p G_N} \partial_t \gsup{d}{\m\n}(x^{\rho}),
\eeq
so in particular the energy density's rate of change, $\partial_t \langle T_{tt}(x^{\rho})\rangle$, is fixed by $\gsup{d}{tt}(x^{\rho})$ alone.

Our goal is to relate $\partial_t \langle T_{tt}(x^{\rho})\rangle$ to $\partial_t \see$, where in the QFT $\see$ is the EE between a sub-region $\mathcal{A}$ and its complement on a Cauchy surface. To compute $\see$ holographically, we consider a codimension-two surface $\cw$ homologous to \(\ca\), with \(\partial \cw = \partial \ca\).  We describe $\cw$'s embedding by a mapping $X^m(\xi)$ from $\cw$'s worldvolume, with coordinates $\xi$, into the background spacetime. We then define $\cw$'s area functional,
\beq
\label{areafunc}
A[\cw] = \int d^{d-1} \xi \,\sqrt{\gamma},
\eeq
where $\gamma$ is the determinant of $\cw$'s worldvolume metric. Extremizing $A$ then gives $\see$~\cite{Hubeny:2007xt,Dong:2016hjy},
\beq
\see = \frac{ A[\cw_{\textrm{ext}}]}{4G_N}.
\eeq

Imagine we have the solution $X^m_{(0)}$ for $\cw_{\textrm{ext}}^{(0)}$'s embedding in a given background geometry $G_{mn}^{(0)}$, which we assume is asymptotically $AdS_{d+1}$, but is otherwise arbitrary. If we perturb the metric, $G_{mn}^{(0)} \to G_{mn}^{(0)} + \delta G_{mn}$, which leads to a change in the embedding, $X^m_{(0)} \to X^m_{(0)} + \delta X^m$, then the change in the EE, $\delta \see$, to leading order in $\delta G_{mn}$ and $\delta X^m$, is~\footnote{We may safely assume that under a small perturbation the topology around the entangling wedge does not change, so the homology constraint does not rule out $\cw^{(0)}_{\textrm{min}}$ in the back-reacted geometry.}
\beq
\label{eq:perturbative_ee}
\delta \see = \frac{1}{4 G_N} \int_{\cw^{(0)}_{\textrm{ext}}} d^{d-1} \xi \, \sqrt{\g} \left( 
\theta_m \d X^m + \frac{1}{2}\Theta^{mn}_{\textrm{ext}} \d G_{mn}
\right),
\eeq
where $\theta_m$ and $\Theta^{mn}_{\textrm{ext}}$ are variations of $A$, evaluated on the unperturbed solutions,
\beq
\theta_m \equiv \left.\frac{1}{\sqrt{\g}}\frac{\d A}{\d X^m}\right\vert_{X^m_{(0)},\, G^{(0)}_{mn}},
\hspace{1cm}
\Theta^{mn}_{\textrm{ext}} = \left.\frac{2}{\sqrt{\g}} \frac{\d A}{\d G_{mn}}\right\vert_{X^m_{(0)},\, G^{(0)}_{mn}}.
\eeq
As argued for example in refs.~\cite{Nozaki:2013vta, Chang:2013mca,Lashkari:2013koa}, because $\cw^{(0)}_{\textrm{ext}}$ is an extremal surface in the unperturbed geometry $G_{mn}^{(0)}$, by definition $\theta_m = 0$. We therefore find
\beq
\label{eq:perturbative_ee_2}
\delta \see = \frac{1}{8 G_N} \int_{\cw^{(0)}_{\textrm{ext}}} d^{d-1}\xi \, \sqrt{\g} \,
\Theta^{mn}_{\textrm{ext}} \, \d G_{mn},
\eeq
which generalizes the result of ref.~\cite{Chang:2013mca} for $\delta \see$ to $t$-dependent perturbations.

Eq.~\eqref{eq:perturbative_ee_2} is valid for any holographic spacetime, but for our proof of a FLOER we impose a few restrictions, as follows. First, we assume $G_{mn}^{(0)}$ is asymptotically $AdS_{d+1}$, and so admits a FG form, and is invariant under translations and rotations in the $\vec{x}$ directions as well as translations in $t$, so that
\beq
G^{(0)}_{mn} dx^mdx^n = \frac{L^2}{z^2} \left(dz^2 +  g_{tt}\,dt^2 + g_{xx}\,d\vec{x}^2 \right),
\eeq
where $g_{tt}$ and $g_{xx}$ depend only on $z$. In our examples in the following sections, $G_{mn}^{(0)}$ will be Poincar\'e patch $AdS_{d+1}$ or an $AdS_{d+1}$ black brane. The assumption that $G_{mn}^{(0)}$ is $t$-independent means the extremal surface $\cw^{(0)}_{\textrm{ext}}$ will actually be a \textit{minimal} surface, $\cw^{(0)}_{\textrm{min}}$, and hence also $\Theta^{mn}_{\textrm{ext}} \to \Theta^{mn}_{\textrm{min}}$, the notation that we will use in the following.

We also make three assumptions about the perturbation $\delta G_{mn}$. First, we assume $\delta G_{mn}$ preserves the $AdS_{d+1}$ FG asymptotics, and also preserves translational and rotational symmetry in $\vec{x}$, so that
\beq
\label{eq:flucFGexp}
\delta G_{mn} dx^mdx^n = \frac{L^2}{z^2}  \left( \delta g_{tt}\,dt^2 + \delta g_{xx}\,d\vec{x}^2 \right),
\eeq
where $g_{tt}$ and $g_{xx}$ depend only on $z$ and $t$. In particular, as mentioned above, in $\delta g_{tt}$ and $\delta g_{xx}$'s FG expansions we assume that the first $t$-dependent coefficients are $\delta g_{tt}^{(d)}$ and $\delta g_{xx}^{(d)}$, respectively. All of these assumptions are crucial for our proof of the FLOER, except for translational and rotational symmetry in $\vec{x}$, which we assume only for simplicity of our presentation, but which could be relaxed without spoiling the FLOER. Moreover, our assumptions are relatively mild, being satisfied by an enormous class of holographic spacetimes.

Under these assumptions, plugging the FG expansion of $\delta G_{mn}$ into eq.~\eqref{eq:perturbative_ee_2} and taking $\partial_t$ of both sides gives us
\beq
\label{eq:rate_of_ee_change}
\partial_t \d\see = \frac{L^2}{8 G_N} \int_{\cw^{(0)}_{\textrm{min}}} d^{d-1}\xi \, \sqrt{\g} \, \Theta^{\m\n}_{\textrm{min}} \, z^{d-2}  \, \partial_t \d \gsup{d}{\m\n}(t)+\ldots,
\eeq
where $\ldots$ indicates higher powers of \(z\), which are suppressed for a sub-region sufficiently small compared to any other scale. We will henceforth assume that the sub-region is sufficiently small to neglect the $\ldots$ terms.

To proceed any further we need an explicit form for $\Theta^{\mu\nu}_{\textrm{min}}$, for which we must restrict to specific $\mathcal{A}$. We will use two different $\mathcal{A}$'s: a sphere, defined by \(\vert\vec{x}\vert\leq R\), and a strip, defined as two parallel planes separated in $x^1 \equiv x$ by a distance $\ell$, and symmetric about $x=0$.

For the sphere, we employ spherical coordinates, with radial coordinate $r$. By spherical symmetry we can then parameterize $\cw$'s embedding as $r(z)$, so that
\beq
\label{eq:induced_metric_det_sphere}
\sqrt{\g}  = \left(\frac{L}{z}\right)^{d-1}r^{d-2} g_{xx}^{(d-2)/2} \sqrt{h} \,\sqrt{1 + g_{xx}\,r'^2},
\eeq
where $h$ is the determinant of the metric $h_{\alpha\beta}$ of a unit $(d-2)$ sphere, $S^{d-2}$. We then find
\beq
\label{eq:ST_sphere}
\Theta^{mn}_{\textrm{min}} \, \partial_m \otimes \partial_n = \left(\frac{z}{L}\right)^2 \left( \frac{(\partial_z + r' \partial_r)^2}{1 + g_{xx}r'^2} + \frac{1}{r^2 g_{xx}} h^{\a\b} \partial_{\a} \otimes \partial_\b \right).
\eeq

For the strip, by translational symmetry in the $\vec{x}$ directions we can parameterize $\cw$'s embedding as $x(z)$, so that
\beq
\label{eq:induced_metric_det_strip}
\sqrt{\g}= \left(\frac{L}{z}\right)^{d-1} g_{xx}^{(d-2)/2} \sqrt{1 + g_{xx}\,x'^2},
\eeq
\beq
\label{eq:ST_strip}
\Theta^{mn}_{\textrm{min}} \, \partial_m \otimes \partial_n=  \left(\frac{z}{L}\right)^2 \left( \frac{(\partial_z + x' \partial_x)^2}{1 + g_{xx}x'^2} + \frac{1}{g_{xx}} \d^{\alpha\beta} \partial_{\alpha} \otimes \partial_\beta \right).
\eeq
Since the $\sqrt{\gamma}$ in eq.~\eqref{eq:induced_metric_det_strip} depends only on $x'(z)$, and not on $x(z)$, if we plug eq.~\eqref{eq:induced_metric_det_strip} into the area functional eq.~\eqref{areafunc}, then variation with respect to $x'(z)$ gives us a constant of motion, $\kappa$. We can then solve algebraically for $x'(z)$ in terms of $\kappa$,
\beq
\label{eq:strip_equation_of_motion}
x'(z) =\pm\frac{1}{\sqrt{\kappa^{d-1}z^{2-2d}g_{xx}^{d} - g_{xx}}},
\hspace{1cm} \kappa = \frac{z_*^2}{g_{xx}^*},
\eeq
where \(z_\star\) denotes $\cw^{(0)}_{\textrm{min}}$'s maximal extension in $z$, fixed by integrating $x'(z)$ from \(z=0\) to \(z_\star\) with the boundary conditions \(x(0)=\pm\ell/2\) and by symmetry \(x(z_\star)=0\), and $g_{xx}^* \equiv g_{xx}(z_*)$.

We now plug the $\Theta^{mn}_{\textrm{min}}$ from eqs.~\eqref{eq:ST_sphere} and~\eqref{eq:ST_strip} into eq.~\eqref{eq:rate_of_ee_change} for $\partial_t \see$. Crucially, the $\Theta^{mn}_{\textrm{min}}$ in eqs.~\eqref{eq:ST_sphere} and~\eqref{eq:ST_strip} depend only on $z$, so we can trivially perform the integration over all other worldvolume coordinates $\xi$. Moreover, in the sum over $\mu$ and $\nu$ in $\Theta^{\m\n}_{\textrm{min}} \, \partial_t \d \gsup{d}{\m\n}(t)$, only the $\vec{x}$ directions contribute, and indeed all contribute equally, due to the rotational symmetry in the $\vec{x}$ directions. Dropping the $\ldots$ terms in eq.~\eqref{eq:rate_of_ee_change}, as mentioned above, we thus find, for the sphere and strip, respectively,
\begin{subequations}
\label{floergxx}
	\begin{align}
	\partial_t \d \see^\mathrm{sphere} 
	&= \frac{L^{d-1}}{8G_N} \vol(S^{d-2}) \partial_t \d \gsup{d}{xx}
	\int_0^{z_*} dz \,
	z r^{d-2} g_{xx}^{\frac{d}{2} - 2} \sqrt{1+ g_{xx}r'^2}
	\left(\frac{g_{xx}r'^2}{1 + g_{xx} r'^2} + d-2
	\right),
	\\
	\partial_t \d\see^\mathrm{strip}
	&= \frac{L^{d-1}}{4G_N} \vol(\mathbb{R}^{d-2}) \partial_t \d\gsup{d}{xx} \int_0^{z_*} dz \, z g_{xx}^{\frac{d}{2}-2} \frac{(g_{xx}^* z^2/g_{xx} z_*^2)^{d-1} + d - 2}{\sqrt{1-(g_{xx}^* z^2/g_{xx} z_*^2)^{d-1}}},
	\end{align}
\end{subequations}
where in both cases $z_*$ denotes $\cw^{(0)}_{\textrm{min}}$'s maximal extension in $z$.

We can write each right-hand-side in eq.~\eqref{floergxx} in terms of $\partial_t E$, with $E$ the energy inside $\mathcal{A}$, as follows. Translational and rotational symmetry in $\vec{x}$ implies $\langle T_{\mu\nu}\rangle$ is $\vec{x}$-independent, so $\partial_t E$ is simply the volume of $\mathcal{A}$ times $\partial_t \langle T_{tt}\rangle$. From eq.~\eqref{eq:energy_momentum_derivative} we have $\partial_t \langle T_{tt} \rangle \propto \partial_t g_{tt}^{(d)}$, however the right-hand-sides of eq.~\eqref{floergxx} involve $\partial_t g_{xx}^{(d)}$. To replace $g_{xx}^{(d)}$ with $g_{tt}^{(d)}$, we use the fact that $T_{\mu\nu}$ is traceless, $T_{\mu}^{~\mu}=0$, up to a possible Weyl anomaly in even $d$, and the fact that the Weyl anomaly is $t$-independent for $G_{mn}^{(0)}$ obeying our assumptions, so that $\partial_t T_{\mu}^{~\mu}=0$ in any $d$. As a result, $\partial_t g_{tt}^{(d)} = (d-1) \partial_t g_{xx}^{(d)}$ in any $d$. Plugging that into eq.~\eqref{eq:energy_momentum_derivative} and multiplying by $\mathcal{A}$'s volume we find (for the sphere, the volume of a $(d-1)$ unit ball is \(\vol(S^{d-2})/(d-1)\)) 
\begin{subequations}
	\begin{align}
	\partial_t E^\mathrm{sphere} 
	&= \frac{d L^{d-1}}{16\p G_N} \, \vol(S^{d-2})\, R^{d-1} \partial_t g^{(d)}_{xx},
	\\
	\partial_t E^\mathrm{strip}
	&= \frac{d L^{d-1}}{16\p G_N} \,\vol(\mathbb{R}^{d-2})\, (d-1)\, \ell \, \partial_t g^{(d)}_{xx}.
	\end{align}
\end{subequations}
From eq.~\eqref{floergxx} we thus identify our FLOER,
\beq
\partial_t \see = \frac{\partial_t E}{\tent},
\eeq
with entanglement temperature $\tent$ for the sphere and strip, respectively,
\begin{subequations}
	\begin{align}
	(\tent^\mathrm{sphere})^{-1} &= \frac{2\p}{d R^{d-1}} \int_0^{z_\star} dz \,
	z r^{d-2} g_{xx}^{\frac{d}{2} - 2} \sqrt{1+ g_{xx}r'^2}
	\left(\frac{g_{xx}r'^2}{1 + g_{xx} r'^2} + d-2
	\right), \label{eq:generaltentsphere}
	\\
	(\tent^\mathrm{strip})^{-1} &= \frac{4\p}{d(d-1)\ell} \int_0^{z_\star} dz \, z
	g_{xx}^{\frac{d}{2}-2} \frac{(g_{xx}^\star z^2/g_{xx} z_\star^2)^{d-1} + d - 2}{\sqrt{1-(g_{xx}^\star z^2/g_{xx} z_\star^2)^{d-1}}}. \label{eq:generaltentstrip}
	\end{align}
\end{subequations}

If $G_{mn}^{(0)}$ is pure $AdS_{d+1}$, where $g_{xx}=1$, then $\cw_{\textrm{min}}^{(0)}$ for the sphere is given by $r(z)  = \sqrt{R^2 - z^2}$, for which $z_*=R$, and for the strip,  \(z_\star = \ell\,\G(\frac{1}{2(d-1)})/2\sqrt{\p}\G(\frac{d}{2(d-1)})\)~\cite{Ryu:2006ef}. In these cases $\tent$ takes the same value as in the FLEE, eqs.~\eqref{spheretent} and~\eqref{striptent}, respectively.

In the following sections we identify examples in which the \textit{bulk} stress-energy tensor, $\mathcal{T}_{mn}$, produces a perturbation $\delta G_{mn}$ obeying all of our assumptions, thus leading to a non-trivial FLOER. Moreover, the FLEE in eq.~\eqref{flee2} is typically violated.

\section{\texorpdfstring{$AdS_4$}{AdS4} Vaidya}
\label{vaidya}

In this section we consider solutions of Einstein-Maxwell theory in $AdS_4$, with bulk action
\beq
S = \frac{1}{16 \pi G_N}\int d^4x \, \sqrt{-\textrm{det}(G_{mn})} \left[R+\frac{6}{L^2} - F^2 \right],
\eeq
with Ricci scalar $R$ and $U(1)$ field strength $F_{mn}$. This theory arises for example as a consistent truncation of eleven-dimensional supergravity on $S^7$~\cite{Chamblin:1999tk,Cvetic:1999xp}. In that case, the dual CFT is the ABJM theory~\cite{Aharony:2008ug}, the $\N=6$ SUSY Chern-Simons-matter theory with gauge group $U(N_c)_k \times U(N_c)_{-k}$, in the limits $N_c \to \infty$ and $N_c \gg k^5$, where the Maxwell gauge field is dual to a conserved current $J^{\mu}$ of a $U(1)$ subgroup of the R-symmetry.

A solution of the Einstein-Maxwell theory in $AdS_4$ that describes a constant external electric field $\mathcal{E}$ in the $x$ direction has Vaidya metric,
\begin{align}
\label{Vaidya}
	d s^2=\frac{L^2}{u^2}\left[-\left(1-m(v)u^3\ri)d v^2-2d ud v+d\vec{x}^2\ri],
\end{align}
with holographic coordinate $u$, with asymptotic $AdS_4$ boundary at $u=0$, null time coordinate $v\equiv t-u$, and $m(v)=2\mathcal{E}^2v$~\cite{Horowitz:2013mia}. The metric in eq.~\eqref{Vaidya} is sourced by a $U(1)$ field strength whose only non-zero components are $F_{xv}=-F_{vx}=\mathcal{E}$, which in the CFT describes an $\mathcal{E}$ that produces $\langle J^x \rangle = \sigma \mathcal{E}$ with conductivity $\sigma = L^2/(4 \pi G_N)$~\cite{Horowitz:2013mia}. In the ABJM example, $\sigma= k^{1/2}N^{3/2}_c/(\pi 3 \sqrt{2})$~\cite{Aharony:2008ug}. The bulk stress-energy tensor's only non-zero component is $\mathcal{T}_{vv} = \mathcal{E}^2 u^2/L^2$, which via $v = t-u$ produces both diagonal components $\mathcal{T}_{tt}$ and $\mathcal{T}_{uu}$ and off-diagonal components $\mathcal{T}_{tu}=\mathcal{T}_{ut}$, all $t$-independent, as advertised in section~\ref{intro}.

The metric in eq.~\eqref{Vaidya} is well-defined only when $m(v)>0$, that is, when $v>0$. In that regime, the metric in eq.~\eqref{Vaidya} describes a black brane geometry with a horizon moving outward, towards the $AdS_4$ boundary, in reaction to $\mathcal{E}$ dumping energy into the system at a constant rate $\partial_t \langle T_{tt}\rangle  = \mathcal{E} \langle J^x\rangle = \sigma \mathcal{E}^2$. We can write the metric in eq.~\eqref{Vaidya} in the form $G_{mn}^{(0)} + \delta G_{mn}$, with $G_{mn}^{(0)}$ the metric of pure $AdS_4$, by switching from $v$ to $t=v+u$:
\begin{subequations}
\label{Vaidya2}
\begin{align}
	G_{mn}^{(0)}dx^m dx^n&=\frac{L^2}{u^2}\le(d u^2-dt^2 + d\vec{x}^2\ri),\\
	\label{Vaidya2fluc}
	\d G_{mn}d x^md x^n&=\frac{L^2}{u^2}\le[2\mathcal{E}^2u^3\le(t-u\ri)\le(d t^2-d td u+d u^2\ri)\ri].
\end{align}
\end{subequations}
However, just to be clear, $G_{mn}^{(0)} + \delta G_{mn}$ is an \textit{exact} solution of the (full, non-linear) Einstein equation, not merely a solution to linear order in $\delta G_{mn}$.

Crucially, $G_{mn}^{(0)}+\delta G_{mn}$ obeys all the assumptions in section~\ref{general}, and hence will obey a FLOER. However, we will also compute $\delta \see$ and $\delta E$ themselves, to show that the FLEE of eq.~\eqref{flee2}, $\delta \see = E/\tent$, is violated.

Eq.~\eqref{eq:perturbative_ee_2} gives us the $\delta \see$ induced by $\mathcal{E}$, to leading order in $\mathcal{E}$,
\begin{align}
\label{deltaSeeV}
	\delta \see = \frac{1}{8 G_N} \int_{\cw^{(0)}} d^{d-1}\xi \, \sqrt{\g} \, \Theta_{\textrm{min}}^{uu} \, \d G_{uu},
\end{align}
where in this example $\g$ and $\Theta_{\textrm{min}}^{mn}$ are the determinant of the induced metric and the stress-tensor, respectively, of the minimal surface $\cw_{\textrm{min}}^{(0)}$ in pure $AdS_4$. Again, just to be clear, eq.~\eqref{deltaSeeV} only captures the leading change in the EE due to $\mathcal{E}$, whereas the metric in eq.~\eqref{Vaidya} is an exact solution of the Einstein equation. For a spherical sub-region, we plug the solution for $\cw_{\textrm{min}}^{(0)}$'s embedding, $r(u) = \sqrt{R^2-u^2}$, into eqs.~\eqref{eq:induced_metric_det_sphere} and~\eqref{eq:ST_sphere} for $\gamma$ and $\Theta_{\textrm{min}}^{mn}$, respectively, and then use $\delta G^{uu}$ from eq.~\eqref{Vaidya2fluc}, to find from eq.~\eqref{deltaSeeV}
\beq
\delta \see = \left(\frac{L^2}{4\pi G_N}\mathcal{E}^2 \right) 2 \pi^2 R \int\limits_0^R d u\,u\le(t-u\ri)\le(1-\frac{u^2}{R^2}\ri) = \ce \langle J^x \rangle \left( \pi R^2 \right) \left( \frac{\pi R}{2}\right) \left( t - \frac{8}{15} \, R\right), \nonumber
\eeq
where in the second equality we used $\langle J^x \rangle = \sigma \mathcal{E}$ with $\sigma = L^2/(4 \pi G_N)$. Using the Ward identity for the energy density $\partial_t \langle T_{tt} \rangle = \mathcal{E} \langle J^x \rangle$ and the area $(\pi R^2)$ of a sphere in two spatial dimensions, we identify $\ce \langle J^x \rangle \left( \pi R^2 \right)=\partial_t E$, and from eq.~\eqref{spheretent} with $d=3$, we identify $\tent = 2/(\pi R)$. We thus find
\beq
\label{vaidyadeltaseesphere}
\delta \see = \frac{\partial_t E}{\tent} \left( t - \frac{8}{15} \, R \right).
\eeq
The analogous calculation for a strip sub-region of width $\ell$ gives
\beq
\label{vaidyadeltaseestrip}
\delta \see = \frac{\partial_t E}{\tent} \left( t - \frac{8}{5\pi} \, \ell \right),
\eeq
where $\tent=4\ell/(\pi^2u_*^2)$ with $u_*=\ell\Gamma(1/4)/2\sqrt{\pi}\Gamma(3/4)$ in $d=3$, in agreement with eq.~\eqref{striptent} with $d=3$. As mentioned above, the metric in eq.~\eqref{Vaidya} is valid only for $v = t - u >0$, so eqs.~\eqref{vaidyadeltaseesphere} and~\eqref{vaidyadeltaseestrip} are valid only for $t>R$ or $t>u_*$, respectively, so that in both cases $\delta \see >0$. Eqs.~\eqref{vaidyadeltaseesphere} and~\eqref{vaidyadeltaseestrip} clearly obey the FLOER, $\partial_t \delta \see = \partial_t E/\tent$, as expected. 

To compute $\delta E$ we switch from the coordinate $u$ in eq.~\eqref{Vaidya2} to the FG coordinate $z$ in eq.~\eqref{eq:fefferman_graham_form}, using $1/u^2=g_{xx}/z^2$. Comparing $G_{tt}$ in the two coordinate systems,
\beq
G_{tt} = \frac{L^2}{z^2}\le(-1+z^3 g_{tt}^{(3)}+\ldots\ri)=\frac{L^2}{u^2}\le(-1+u^3\le(g_{tt}^{(3)}+g_{xx}^{(3)}\ri)+\ldots\ri),
\eeq
we find $g_{tt}^{(3)}+g_{xx}^{(3)} = 2 \mathcal{E}^2 t$. Tracelessness of $T_{\mu\nu}$ gives us $g_{xx}^{(3)}=g_{tt}^{(3)}/2$, so that $g_{tt}^{(3)}=4\mathcal{E}^2t/3$. Eq.~\eqref{eq:general_energy_momentum} then gives the energy density,
\beq
\langle T_{tt} \rangle =\frac{3L^2}{16\pi G_N} g_{tt}^{(3)}=\frac{L^2}{4\pi G_N} \, \mathcal{E}^2 \, t  = \mathcal{E} \langle J^x \rangle t,
\eeq
so that, unsurprisingly, $\partial_t \langle T_{tt} \rangle = \mathcal{E} \langle J^x \rangle$. As a result, for spherical and strip sub-regions, $\delta E = \mathcal{E} \langle J^x \rangle (\pi R^2) \, t$ and $\delta E = \mathcal{E} \langle J^x \rangle (\ell \,\textrm{Vol}(\mathbb{R})) t$, respectively, or more simply, $\delta E = t \,\partial_tE$.

For perturbations of the CFT state, without changes to the CFT Hamiltonian, intuition from QFT~\cite{Blanco:2013joa} and results from holography~\cite{Bhattacharya:2012mi} suggest that for a sub-region of fixed size the FLEE of eq.~\eqref{flee2} should hold for sufficiently small $\delta E$. Strictly speaking, that criterion does not immediately translate to our case, because we deform the CFT Hamiltonian, by $\mathcal{E}$. Nevertheless, na\"ively applying that criterion to our case, we expect the FLEE to hold for $t$ short enough that $\mathcal{E}$ has deposited little energy into the sub-region. For example for the sphere we expect the FLEE to hold for $t$ short enough that $\delta E =t \, \partial_t E \lesssim 1/R$, meaning $t \lesssim \left(\mathcal{E} \langle J^x \rangle \pi R^3\right)^{-1}$. We can make that time arbitrarily long by making $\mathcal{E}$ arbitrarily small. In particular, the times for which we expect the FLEE to hold can be made $\gg R$, and hence can easily include the regime $t > R$ where our result for $\delta \see$ eq.~\eqref{vaidyadeltaseesphere} is valid. However, plugging a time of order $R$ into $\delta \see$ in eq.~\eqref{vaidyadeltaseesphere}, we find that $\delta \see \neq \delta E / \tent$, due to the term $\propto R$ in eq.~\eqref{vaidyadeltaseesphere}. Of course analogous statements apply for $\delta \see$ of the strip in eq.~\eqref{vaidyadeltaseestrip}. In short, in both cases we find that the FLEE of eq.~\eqref{flee2} is violated, as advertised.

Moreover, as mentioned in section~\ref{intro}, the ``entanglement tsunami'' model~\cite{Liu:2013iza,Liu:2013qca,Casini:2015zua} offers a possible explanation for the offending terms, as a difference in initial conditions. As soon as $\mathcal{E}$ is turned on, it pumps energy into the CFT and begins producing massless EPR pairs, doing both at a constant rate and uniformly throughout space. However, the pairs produced at sufficiently early times only contribute to EE after some finite time required to exit the sub-region $\mathcal{A}$. As a result, $\delta \see$ lags behind $\delta E$ by an amount on the order of $\mathcal{A}$'s size, $R$ or $\ell$, as indeed observed in eqs.~\eqref{vaidyadeltaseesphere} and~\eqref{vaidyadeltaseestrip}. Of course, not all EPR partners are equidistant from $\partial \mathcal{A}$, so the lag is not identically $R$ or $\ell$, but is only $\propto R$ or $\ell$.

\section{D3/D7 with Electric Field}
\label{d3d7}

In this section we study the D3/D7 system~\cite{Karch:2002sh}. Type IIB supergravity in the near-horizon geometry of $N_c \to \infty$ D3-branes, $AdS_5 \times S^5$, is dual to $\N=4$ SYM with gauge group $SU(N_c)$, in the limits $N_c \to \infty$ and 't Hooft coupling $\lambda \to \infty$~\cite{Maldacena:1997re}. A number $N_f$ of probe D7-branes along $AdS_5 \times S^3$ is dual to a number $N_f \ll N_c$ of massless $\N=2$ SUSY hypermultiplets in the fundamental representation of $SU(N_c)$, \textit{i.e.}\ flavor fields~\cite{Karch:2002sh}. The D7-brane worldvolume $U(N_f)$ gauge fields are dual to conserved $U(N_f)$ flavor symmetry currents.

As mentioned in sec.~\ref{intro}, the probe D7-brane provides a \textit{time-independent} example in which the FLEE of eq.~\eqref{flee1} can hold while that in eq.~\eqref{flee2} is violated. Suppose we give the flavor fields a non-zero $\N=2$ SUSY-preserving mass, $m$. The proof of ref.~\cite{Casini:2016udt} applies in that case, so if $\rho$ and $\rho'$ are the vacua of the $m=0$ and $m\neq 0$ theories, then we expect  $S(\rho|\rho')\geq 0$ and hence the FLEE of eq.~\eqref{flee1}. For the FLEE of eq.~\eqref{flee2}, SUSY guarantees $\delta E = 0$. On the other hand, holographic results for $\delta \see$ of a spherical sub-region~\cite{Kontoudi:2013rla,Karch:2014ufa} include a term $\propto (mR)^2 \log (\e/R)$, with FG cut-off $\e$. The coefficient of the the $\log(\e/R)$ cannot be set to zero by re-scaling $\e$, so clearly $\delta S \neq \delta E/\tent$, \textit{i.e.}\ the FLEE of eq.~\eqref{flee2} is violated.

To realize out \textit{time-dependent} example, we introduce $T \neq 0$, so that $AdS_5$ becomes an $AdS_5$ black brane. The $\N=4$ SYM and flavor contributions to $\langle T_{\mu\nu}\rangle$ are then order $N_c^2$ and $N_f N_c\ll N_c^2$, respectively~\cite{Karch:2008uy}, so we may think of the flavors as probes inside an enormous heat bath. We also introduce a constant, external electric field $\mathcal{E}$ in the $x$ direction for the diagonal $U(1)\subset U(N_f)$, producing a current, $\langle J^x \rangle$, of charge carriers in the flavor sector. The charge density vanishes, $\langle J^t \rangle=0$, so the current comes entirely from Schwinger pair production~\cite{Hashimoto:2013mua,Hashimoto:2014dza}. We consider NESS in which $\langle J^x \rangle$ is $t$-independent because the charge carriers gain energy from $\mathcal{E}$ at the same constant rate $\mathcal{E} \langle J^x \rangle$ that they lose energy to the heat bath~\cite{Karch:2008uy}, as we discuss below.

We use an $AdS_5$ black brane metric
\beq
\label{eq:ads_schwarzschild}
ds^2 = \frac{L^2}{u^2} \left(\frac{du^2}{b(u)} - b(u) dt^2 + d\vec{x}^2 \right), \qquad b(u) = 1 - (u/u_h)^4,
\eeq
with $T = 1/(\pi u_h)$. The D7-branes fill the $AdS_5$ black brane space and also wrap an equatorial $S^3 \subset S^5$ with radius $L$. The only non-trivial contribution to the D7-brane action, $S_{D7}$, is then the DBI term,
\beq
\label{d7action}
S_\mathrm{D7} = - N_f T_\mathrm{D7} \int d^8 \zeta \sqrt{-\det(\Gamma_{ab} + (2\pi \alpha') F_{ab})},	
\eeq
with D7-brane tension $T_{D7}=(2\pi)^{-7}\alpha'^{-4}g_s^{-1}$, with string length squared $\alpha'$ and coupling $g_s$, worldvolume coordinates $\zeta^a$ with $a=0,\ldots,7$, worldvolume metric $\Gamma_{ab}$, and worldvolume $U(1)$ field strength $F_{ab}= \partial_a A_b -\partial_b A_a$. To describe $\mathcal{E}$ and $\langle J^x \rangle$ we make the ansatz
\beq
\label{d7ansatz}
A_x(t,u) = - \ce t + a_x(u),
\eeq
with all of $A_a$'s other components zero. Plugging our ansatz eq.~\eqref{d7ansatz} into $S_{D7}$ in eq.~\eqref{d7action}, and trivially performing the integration over the $S^3$ directions, we find
\beq
\label{d7actionansatz}
S_\mathrm{D7} = - N_f T_\mathrm{D7} L^3 \vol(S^3) \int d^5 \zeta \, \frac{L^5}{u^5} \sqrt{1 + (2\pi\alpha')^2\frac{u^4}{L^4} \left( b(u) a_x'^2(u) - \frac{\ce^2}{b(u)}\right)}.
\eeq
For simplicity, we define an ``effective tension'',
\beq
\tilde{T}_{D7} \equiv N_f T_\mathrm{D7} L^3 \vol(S^3) = \frac{\lambda N_f N_c}{\left(2\pi\right)^4} \frac{1}{L^5},
\eeq
where in the second equality we used $\vol(S^3)=2\pi^2$, $\lambda \equiv 4 \pi g_s N_c$, and $\lambda = L^4/\alpha'^2$~\cite{Karch:2007pd}.

Crucially, $S_{D7}$ in eq.~\eqref{d7actionansatz} depends on $a_x'(u)$ but not on $a_x(u)$, hence we have a first integral of motion, which in the dual CFT is precisely the current: $\frac{\delta S_{D7}}{\delta a_x'} = \langle J^x \rangle$~\cite{Karch:2007pd}. We can then solve algebraically for $a_x'(u)$ in terms of $\langle J^x \rangle$,
\beq
\label{axsol}
a_x'(u) = \frac{\langle J^x \rangle}{b(u)L} \sqrt{\frac{b(u)/u^4 - (2\pi\alpha')^2\ce^2/ L^4}{\tilde{T}_{D7}^2 (2\pi\alpha')^2 b(u)/u^6 - \langle J^x \rangle^2/L^6}}.
\eeq
To fix $\langle J^x \rangle$ we follow ref.~\cite{Karch:2007pd}: we plug the solution for $a_x'(u)$ in eq.~\eqref{axsol} into $S_{D7}$ in eq.~\eqref{d7actionansatz} and demand that the result remains real for all \(u \in [0, u_h]\), since a non-zero imaginary part of an effective action signals a tachyon~\cite{Hashimoto:2013mua,Hashimoto:2014dza}. We find $\langle J^x \rangle = \sigma \mathcal{E}$, with conductivity
\beq
\sigma = \frac{N_fN_c T}{4 \pi} \left[1+\mathcal{E}^2/\left(\pi \sqrt{\lambda}\,T^2/2\right)^2\right]^{1/4}.
\eeq

To compute the $\delta \see$ due to $\mathcal{E}$, we must compute the perturbative back-reaction of the D7-branes to first order. At first, that looks like a daunting task, since the D7-branes couple not only to the metric but also to the axio-dilaton and $B$-field, and moreover break several symmetries of the background, for example breaking the $S^5$'s $SO(6)$ isometry down to the $SO(4)\times U(1)$ preserved by the equatorial $S^3 \subset S^5$. Fortunately, however, as argued in refs.~\cite{Chang:2013mca,Chang:2014oia}, if the D7-brane worldvolume fields are independent of the $S^3\subset S^5$ directions, as in our case, then using an ``effective stress-energy tensor'', obtained by integrating the $AdS_5$ part of the D7-brane stress energy tensor over the $S^3$, is sufficient for computing $\delta \see$. In our case, this effective stress-energy tensor is
\beq
\mathcal{T}_\mathrm{eff}^{mn} = - \tilde{T}_{D7} \frac{\sqrt{-\det(\Gamma_{mn}+(2\pi\alpha')F_{mn})}}{\sqrt{-\det(G_{mn}^{(0)})}} \left[(\Gamma + (2\pi\alpha')F)^{-1}\right]^{(mn)},
\eeq
where $G_{mn}^{(0)}$ is the $AdS_5$ black brane metric in eq.~\eqref{eq:ads_schwarzschild}, $\Gamma_{mn}$ and $F_{mn}$ are now restricted to the directions in eq.~\eqref{eq:ads_schwarzschild}, and $(mn)$ indicates symmetrization over the indices $m$ and $n$. Splitting $\mathcal{T}_\mathrm{eff}^{mn}$ into diagonal and off-diagonal parts, $\mathcal{T}^\mathrm{eff}_{mn} = \mathcal{T}^\mathrm{diag}_{mn} + \mathcal{T}^\mathrm{off}_{mn}$, for the $a_x'(u)$ solution in eq.~\eqref{axsol} we find
\begin{subequations}
\label{eq:d3d7_energy_momentum}
	\begin{align}
	\mathcal{T}^\mathrm{diag}_{mn}dx^mdx^n = - \frac{ a_x' L^3}{\langle J^x \rangle u^3} &\left[ \le(1 - \frac{\langle J^x \rangle^2}{\tilde{T}_{D7}^2(2\pi\alpha')^2}\frac{u^6}{b L^6}\ri)du^2
	-b \frac{b^2 - \langle J^x\rangle^2 \ce^2 u^{10}/(\tilde{T}_{D7}^2L^{10})}{b -(2\pi\alpha')^2\ce^2 u^4/L^4}\,dt^2\ri.\nn\\
	&\quad\quad\le.+
	\frac{\langle J^x \rangle^2 u^2}{\tilde{T}_{D7}^2\,b\, a_x'^2 L^2}\,dx^2+b \,(dx^2)^2+
	b \,(dx^3)^2
	\right],
	\\
	\mathcal{T}_{mn}^\mathrm{off}dx^mdx^n &= - \ce \langle J^x \rangle \frac{u^3}{b(u)L^3} \, 2 du \, dt.
	\end{align}
\end{subequations}
As advertised in section~\ref{intro}, $\mathcal{T}^{mn}_{\textrm{eff}}$ is $t$-independent but has off-diagonal terms $\mathcal{T}^{ut}_{\textrm{off}}=\mathcal{T}^{tu}_{\textrm{off}}$. In fact, $\mathcal{T}^\mathrm{diag}_{mn}$ and $\mathcal{T}^\mathrm{off}_{mn}$ turn out to be separately conserved, so if we linearize Einstein's equation in $\delta G_{mn}$, and split $\delta G_{mn}$ into parts sourced by $\mathcal{T}^\mathrm{diag}_{mn}$ and $\mathcal{T}^\mathrm{off}_{mn}$, respectively, $\delta G_{mn}= \delta G_{mn}^{\textrm{diag}}+\delta G_{mn}^{\textrm{off}}$ (which are not necessarily diagonal and off-diagonal themselves), then we can solve for $\delta G_{mn}^{\textrm{diag}}$ and $\delta G_{mn}^{\textrm{off}}$ separately.

We have checked explicitly that a $t$-independent solution for $\delta G_{mn}^{\textrm{diag}}$ exists, whose existence relies crucially on the fact that $\mathcal{T}^\mathrm{diag}_{mn}$ is invariant under $t$-reversal. At leading order in $\mathcal{E}$, $\mathcal{T}^\mathrm{diag}_{mn}$'s back-reaction is just a shift of the cosmological constant, as expected: the DBI action in eq.~\eqref{d7action} with trivial worldvolume fields is a contribution to the cosmological constant $\propto T_{D7}$. The cosmological constant is $\propto 1/L^2$, and roughly speaking $L$ in Planck units is dual to the number of degrees of freedom in the CFT, measured for example in even $d$ by a central charge~\cite{Freedman:1999gp}. In particular, adding a space-filling probe DBI action with trivial worldvolume fields corresponds to adding degrees of freedom, such as adding flavor fields to $\N=4$ SYM. Such a deformation results in a FLEE of the form in eq.~\eqref{flee2}, but with a ``chemical potential'' term arising from the change in the number of degrees of freedom~\cite{Kastor:2014dra}.

On the other hand, $\mathcal{T}^\mathrm{off}_{mn}$ breaks $t$-reversal, and hence so does $\delta G_{mn}^{\textrm{off}}$. Indeed, the solution for $\delta G_{mn}^{\textrm{off}}$ is
\beq
\label{D7offdiagsol}
\d G^\mathrm{off}_{mn} dx^mdx^n = \frac{16\pi G_N }{3 L} \, \ce \langle J^x \rangle \, t \, u^2 \left(dt^2 + \frac{du^2}{b^2(u)} \right).
\eeq
If $\d G^\mathrm{off}_{mn}$ grows too big, then the linearized approximation breaks down, hence the linearized solution in eq.~\eqref{D7offdiagsol} is valid only for sufficiently small $\ce \langle J^x \rangle t$.

Strictly speaking, in this example $G_{mn}=G_{mn}^{(0)} + \delta G_{mn}$ does not obey all the assumptions in section~\ref{general}. For instance, as mentioned above $\delta G_{mn}^{\textrm{diag}}$ is asymptotically $AdS_5$, but shifts $L$, something we did not account for in section~\ref{general}. However, a key step in section~\ref{general} was taking $\partial_t$ of $\delta \see$, so in fact we only need the $t$-dependent part of $\delta G_{mn}$ to obey our assumptions. In this example, all of $\delta G_{mn}$'s $t$-dependence is in $\delta G_{mn}^{\textrm{off}}$, and indeed, $G_{mn}^{(0)}+G_{mn}^{\textrm{off}}$ obeys all the assumptions in section~\ref{general}, and hence this example must obey a FLOER.

However, to dispel any doubt, we have calculated $\partial_t \see$ following the steps in section~\ref{general}, adapted to the coordinate $u$ of eq.~\eqref{eq:ads_schwarzschild}, with the results
\begin{subequations}
	\begin{align}
	\partial_t \d \see^\mathrm{sphere} &= \mathcal{E} \langle J^x \rangle \left(\frac{4}{3}\pi R^3\right)\frac{2\pi}{R^3}\int_0^{u_\star}du \frac{ r^2 u}{\sqrt{b^3(u) (1 + b(u) r'^2)}},
	\\
	\partial_t \d \see^\mathrm{strip} &= \mathcal{E} \langle J^x \rangle\left( \ell \, \vol(\mathbb{R}^2)\right) \frac{4\pi}{3\ell} \int_0^{u_\star} du \, u \sqrt{\frac{1-(u/u_\star)^6}{b^3(u)}},
	\end{align}
\end{subequations}
where for the sphere $r(u)$ is the solution for the the minimal surface's embedding, and for both the sphere and strip \(u_\star\) is the minimal surface's maximal extension in $u$, in the unperturbed $AdS_5$ black brane geometry of eq.~\eqref{eq:ads_schwarzschild}. For the strip, $u_\star$ is related to the width $\ell$ by
\beq
\ell = 2 \int_0^{u_\star} du \frac{(u/u_\star)^3}{\sqrt{b(u) (1 -(u/u_\star)^6)}}.
\eeq
Identifying $\partial_t \delta E=\ce \langle J^x \rangle \left(\frac{4}{3}\pi R^3\right)$ or $\partial_t \delta E=\ce \langle J^x \rangle \left( \ell \, \vol(\mathbb{R}^2)\right)$ for the sphere and strip, respectively, we thus find
\begin{subequations}
\label{eq:d3d7_first_law}
	\begin{align}
	\partial_t \d \see^\mathrm{sphere}
	&= \partial_t \d E \,
	\frac{2 \p}{R^3} \int_0^{u_\star} du \frac{r^2 u}{\sqrt{b^3(u)(1 + b(u) r'^2)}},  \label{eq:d3d7_first_law_sphere}
	\\
	\partial_t \d \see^\mathrm{strip} 
	&= \partial_t \d E \,  
	\frac{4 \p}{3 \ell} \int_0^{u_\star} du \,  u \sqrt{\frac{1-(u/u_\star)^6}{b^3(u)}} \label{eq:d3d7_first_law_strip}.
	\end{align}
\end{subequations}
A straightforward calculation confirms that for the $AdS_5$ black brane the integrals in eqs.~\eqref{eq:d3d7_first_law_sphere} and~\eqref{eq:d3d7_first_law_strip} reproduce $\tent$ from eqs.~\eqref{eq:generaltentsphere} and~\eqref{eq:generaltentstrip}, respectively. We have thus explicitly shown that this example obeys the FLOER.

As mentioned in section~\ref{intro}, the FLOER may be useful because $\partial_t \delta E$ is often easier to calculate than $\partial_t\delta \see$. Indeed, for probe branes we can calculate $\partial_t \delta E$ in the probe limit, without computing back-reaction, following refs.~\cite{Karch:2008uy,Das:2010yw}. The probe flavor's order $N_f N_c$ contribution to the energy density, $\delta \langle T_{tt} \rangle$, is given holographically by the energy density on the D7-brane, $\mathcal{T}^t_{~t}$, integrated over the $S^3\subset S^5$ and $u$ directions,
\beq
\label{eq:D7totalenergy}
\delta \langle T_{tt} \rangle = - \int_0^{u_h} du \sqrt{-\textrm{det}(\Gamma_{ab})} \, \,{\mathcal{T}^t}_t.
\eeq
Taking $\partial_t$ of eq.~\eqref{eq:D7totalenergy} and using \(\nabla_c (\sqrt{-\textrm{det}(\Gamma_{ab})} \, {\mathcal{T}^c}_t) = 0\), from conservation of $\mathcal{T}_{ab}$, we find
\beq
\label{eq:D7totalrate}
\partial_t \delta \langle T_{tt} \rangle = \int_0^{u_h} du \, \partial_u \sqrt{-\textrm{det}(\Gamma_{ab})} \, {\mathcal{T}^u}_t 
= \left[\sqrt{-\textrm{det}(\Gamma_{ab})} \, {\mathcal{T}^u}_t \right]^{u_h}_0.
\eeq
From the $\mathcal{T}^u_{~t}$ in eq.~\eqref{eq:d3d7_energy_momentum}, we find that the rate of energy density gain at the boundary, dual to the energy density pumped into the probe sector by $\mathcal{E}$, and energy density loss at the horizon, dual to the energy density the probe sector dumps into the heat bath, are equal:
\beq
\left.\sqrt{-\textrm{det}(\Gamma_{ab})} \, {\mathcal{T}^u}_t\right\vert_{u=0}
	=\left.\sqrt{-\textrm{det}(\Gamma_{ab})} \, {\mathcal{T}^u}_t\right\vert_{u=u_h} 
	= -\mathcal{E}\langle J^x \rangle.
\eeq
The \textit{total} rate of change of energy density in eq.~\eqref{eq:D7totalrate} thus vanishes, $\partial_t \delta \langle T_{tt} \rangle=0$, producing a NESS, as advertised. Presumably, the $\partial_t \delta E$ that appears in the FLOER in eq.~\eqref{eq:d3d7_first_law} comes from the energy injected into the sub-region by $\mathcal{E}$, \textit{i.e.}\ from the $u=0$ contribution to $\partial_t \delta \langle T_{tt} \rangle$ in eq.~\eqref{eq:D7totalrate}. In short, we can calculate $\partial_t \delta E$ directly in the probe limit, avoiding any back-reaction, simply by evaluating $\sqrt{-\textrm{det}(\Gamma_{ab})} \, {\mathcal{T}^u}_t$ at $u=0$.

In general, when $\mathcal{E}\neq0$ a probe brane's induced metric $\Gamma_{ab}$ has a horizon distinct from that of the background metric~\cite{Karch:2007pd,Erdmenger:2007bn,Albash:2007bq}. A temperature can be associated with the worldvolume horizon~\cite{Filev:2008xt,Kim:2011qh,Sonner:2012if,Nakamura:2013yqa,Kundu:2013eba,Kundu:2015qda,Banerjee:2015cvy,Banerjee:2016qeu}, which in general is distinct from the background temperature $T$, clearly indicating that the system is out of equilibrium. The worldvolume horizon may represent the EE of the Schwinger pairs produced by $\mathcal{E}$~\cite{Sonner:2013mba}. However, whether any meaningful notion of entropy can be associated to the worldvolume horizon is unclear. An obvious guess is a Bekenstein-Hawking entropy, the horizon's area over $4 G_N$. However, the DBI action does not describe gravitational degrees of freedom, and $\Gamma_{ab}$ is not necessarily a solution of Einstein's equation, so although we can compute the area of the worldvolume horizon, what should play the role of $4G_N$? The open string coupling~\cite{Kundu:2013eba}? In any case, the worldvolume horizon did not appear to play any special role in our calculation of EE, and in particular, our result for the EE does not appear to be proportional to the area of the worldvolume horizon.

Although we focused on the D3/D7 system, the analysis in this section should straightforwardly generalise to many other systems involving a space-filling probe DBI action with $\mathcal{E} \neq 0$ in an asymptotically $AdS_{d+1}$ spacetime.

\section{Massless Scalars}
\label{massless}

In this section we study holographic CFTs deformed by marginal scalar operators $\mathcal{O}$ with a source linear in time $t$ or in a spatial direction $x$. Explicit examples of such CFTs are $\N=4$ SYM in $d=4$, which has three exactly marginal scalar operators~\cite{Aharony:2002tp,Aharony:2002hx}, and ABJM theory, where the Chern-Simons level, or equivalently the 't Hooft coupling, is exactly marginal.

A marginal scalar operator $\mathcal{O}$ is holographically dual to a massless scalar field $\phi$, whose stress-energy tensor $\mathcal{T}_{mn}$ depends only on derivatives of $\phi$, due to invariance under constant shifts of $\phi$. A linear source for $\mathcal{O}$ produces a $\mathcal{T}_{mn}$ that depends only on the holographic radial coordinate, but may have non-trivial off-diagonal components, producing a $\delta G_{mn}$ that may depend on $t$ or $x$, but obeys the assumptions in section~\ref{general}, hence the FLOER will be obeyed.

However, we compute $\delta \see$ and $\delta E$ separately for $d=2,3,4$, and show that in all cases the FLEE of eq.~\eqref{flee2} is violated. More specifically, we solve Einstein's equation for $\delta G_{mn}$ near the asymptotic $AdS_{d+1}$ boundary, obtaining explicit expressions for only a subset of $\delta G_{mn}$'s FG coefficients, while any remaining FG coefficients could in principle be fixed by imposing regularity of $\delta G_{mn}$ in the bulk. These asymptotic solutions for $\delta G_{mn}$ suffice to establish violation of the FLEE of eq.~\eqref{flee2}.

\subsection{Linear Time Dependence}
\label{lineartime}

In this sub-section we consider $(d+1)$-dimensional Einstein-Hilbert gravity coupled to a massless scalar field $\phi$, with bulk action
\beq
\label{eq:masslessaction}
S = \int d^{d+1}x \, \sqrt{-\textrm{det}(G_{mn})} \left[\frac{1}{16 \pi G_N}\left(R+\frac{d(d-1)}{L^2}\right) - \frac{1}{2}(\partial \phi)^2 \right].
\eeq
As in section~\ref{general}, we consider $G_{mn}=G_{mn}^{(0)}+\delta G_{mn}$ in FG form, with $G_{mn}^{(0)}$ the $AdS_{d+1}$ metric,
\begin{align}
	G_{mn}^{(0)}dx^m dx^n=\frac{L^2}{z^2}\le(dz^2-dt^2+d\vec{x}^2\ri).
\end{align}
A solution for $\f$ admits the FG expansion
\beq
\label{eq:phiFGexp}
\f=\f_0+\dots z^d\f_d+\dots,
\eeq
where the coefficients $\phi_0$, $\phi_d$, etc. generically depend on $t$ and $\vec{x}$. The coefficient $\phi_0$ is dual to the source for $\mathcal{O}$, so we introduce $\phi_0 = -c\,t$ with constant $c>0$ of dimension $[t]^{-1}$. The remaining coefficients in $\f$'s and $\delta G_{mn}$'s FG expansions can then depend only on $t$, although their explicit solutions depend on $d$, so in the following we consider $d=2,3,4$ in turn.

For each of $d=2,3,4$, we compute $\delta \see$ and $\delta E$ for a sphere or strip sub-region. More specifically, to compute $\delta \see$ we use eq.~\eqref{eq:perturbative_ee_2}, whose inputs are $\sqrt{\gamma}$, $\Theta_{\textrm{min}}^{mn}$, and $\delta G_{mn}$. We plug the solution for $\cw_{\textrm{min}}^{(0)}$'s embedding, for example $r(z) = \sqrt{R^2-z^2}$ for the sphere, into eqs.~\eqref{eq:induced_metric_det_sphere} and~\eqref{eq:ST_sphere} to obtain $\gamma$ and $\Theta_{\textrm{min}}^{mn}$, respectively. As mentioned above, we solve for $\delta G_{mn}$ only near the asymptotic $AdS_{d+1}$ boundary, and then extract $\delta E$ via holographic renormalisation~\cite{deHaro:2000xn}. The details of the holographic renormalisation appear in appendix~\ref{holoRG}, where we also check several Ward identities. (In each case, the $\partial_t \langle T_{tt} \rangle$ from holographic renormalisation reproduces eq.~\eqref{eq:energy_momentum_derivative}.) For each of $d=2,3,4$, we find that the FLOER is obeyed, as expected, while the FLEE of eq.~\eqref{flee2} is violated.

\paragraph{Boundary Dimension $d=2$:}

The holographic renormalisation for a massless scalar in $AdS_3$ appears in ref.~\cite{deHaro:2000xn}. Plugging a Minkowski metric at the $AdS_3$ boundary and $\phi_0 = -ct$ into the results of ref.~\cite{deHaro:2000xn} yields
\begin{subequations}
\label{eq:ads3holorenresults}
\begin{align}
	G_{xx}&=\frac{L^2}{z^2}\le[1+z^2 \,g_{xx}^{(2)}-z^2\,\log\le(z^2/L^2\ri)\, 2\pi \frac{G_N}{L}\,c^2+\dots\ri],\\
	\le<T_{xx}\ri>&=\frac{L}{8\pi G_N} g_{xx}^{(2)}+c^2\le(\frac{1}{4}-\eta\ri),\label{eq:ads3holorenresultstxx}\\
	\le<T^\m{}_\m\ri>&=\frac{1}{2}c^2,\label{eq:ads3holorenresultstrace}
\end{align}
\end{subequations}
where the term $\propto \eta$ in $\le<T_{xx}\ri>$ is scheme-dependent, and comes from the finite counterterm
\begin{align}
	S_{CT} = \eta\int d^2x\sqrt{-\textrm{det}(\tilde{g}_{\mu\nu})}\,\,\tilde{g}^{\m\n}\pa_\m\f\pa_\n\f\;,
\end{align}
added to the bulk action $S$ in eq.~\eqref{eq:masslessaction} (with $d=2$) at a regulating cut-off surface $z = \e$, with induced metric $\tilde{g}_{\m\n}$. Plugging $\langle T_{xx}\rangle$ from eq.~\eqref{eq:ads3holorenresultstxx} into $\le<T^\m{}_\m\ri>=\frac{1}{2}c^2$ from eq.~\eqref{eq:ads3holorenresultstrace} then gives $\langle T_{tt} \rangle  = \langle T_{xx} \rangle - \frac{1}{2} c^2$.

The bulk stress-energy tensor $\mathcal{T}_{mn}$ is quadratic in $\partial_m \phi$ and hence $\propto c^2$. We treat $\mathcal{T}_{mn}$ as a perturbation, and so linearize Einstein's equation, producing $\delta G_{mn}$ of order $c^2$. The change in energy inside the sphere $\le|x\ri|<R$
\beq
\delta E= (2R) \delta\langle T_{tt} \rangle = (2R)\frac{L}{8\pi G_N}\delta g_{xx}^{(2)}-(2R)c^2\le(\frac{1}{4}+\eta\ri),
\eeq
is then $\propto c^2$, and in particular, $\delta g_{xx}^{(2)} \propto c^2$. As mentioned above, we compute $\delta \see$ from eq.~\eqref{eq:perturbative_ee_2}, with the result
\beq
\label{eq:masslessd=2}
\delta \see = \frac{\delta E}{\tent} +\frac{\pi}{9}\,c^2\,R^2\le(-6\log\le(2R/L\ri)+8+12\,\eta\ri),
\eeq
where $\tent=3/(2 \pi R)$, as in eq.~\eqref{spheretent} with $d=2$. The $\delta \see$ in eq.~\eqref{eq:masslessd=2} has some scheme dependence, via the term $\propto \eta$, and in particular, a shift of $\eta$ produces a shift of the argument of the logarithm in the term $\propto -R^2\log(2R/L)$. Crucially, however, the choice of $\eta$ is part of the definition of the QFT, so $\eta$ cannot depend on the size $R$ of some arbitrarily-chosen sub-region, and so $\eta$ cannot affect the \textit{coefficient} of the term $\propto -R^2\log(2R/L)$. As a result, the latter coefficient is scheme-independent and hence physically meaningful.

As discussed in section~\ref{vaidya}, we na\"ively expect the FLEE of eq.~\eqref{flee2} to hold at sufficiently early times such that $t \,\partial_t E \lesssim 1/R$. The diffeomorphism Ward identity $\nabla^{\mu} T_{\mu\nu} =  \mathcal{O}\partial_{\nu} \phi_0$ implies $\partial_t \langle T_{tt} \rangle = c \langle \mathcal{O} \rangle$, and hence $\partial_t E = (2R) c \langle \mathcal{O} \rangle$. We thus expect the FLEE to hold for $t \lesssim \left(c\langle \mathcal{O}\rangle 2R^2\right)^{-1}$, which can be made arbitrarily long by making $c$ arbitrarily small, and can hence include the regime $t \simeq R$. As argued above, $\delta E$'s only dependence on $c$ is $\delta E \propto c^2$, and when $t \simeq R$ dimensional analysis requires $\delta E \propto c^2 R$ or $c^2 R \log R$. In that case, in the $\delta \see$ in eq.~\eqref{eq:masslessd=2} the terms $\propto \delta E/\tent$ and $\propto c^2 R^2$ are of the same order, so the FLEE of eq.~\eqref{flee2} is clearly violated, as advertised.

\paragraph{Boundary Dimension $d=3$:}

The details of the holographic renormalisation for a massless scalar in asymptotically $AdS_4$ spacetimes appear in the appendix. In particular, plugging a Minkowski metric at the $AdS_4$ boundary and $\phi_0 = -ct$ into eqs.~\eqref{eq:d=3_relations} and~\eqref{eq:d=3SET} yields
\begin{subequations}
\begin{align}
	G_{xx}&=\frac{L^2}{z^2}\le[1+z^2\,2\pi \frac{G_N}{L^2}\,c^2+z^3 g_{xx}^{(3)}+\dots\ri],\\
	\le<T_{\mu\nu}\ri>&=\frac{3L^2}{16\pi G_N} g_{\m\n}^{(3)},
\end{align}
\end{subequations}
and $T^{\mu}_{~\mu}=0$, as expected in $d=3$. As in the $d=2$ case above, a linearized perturbation $\delta G_{mn}$ is $\propto c^2$, so the change in the energy inside the sphere $\le|\vec{x}\ri|<R$ is
\begin{align}
	\d E=(\pi R^2)\delta\le<T_{tt}\ri>=(\pi R^2)\frac{3L^2}{16 \pi G_N}\delta g_{tt}^{(3)},
\end{align}
where $\delta g_{tt}^{(3)}\propto c^2$. The $\delta E$ for the strip is identical, but with $(\pi R^2) \to \ell \, \textrm{Vol}(\mathbb{R})$. As mentioned above, from eq.~\eqref{eq:perturbative_ee_2} we compute $\delta \see$ for the sphere,
\beq
\label{eq:masslessd=3sphere}
\delta \see =  \frac{\delta E}{\tent}+\frac{2\pi}{3}c^2\,R^2,
\eeq
where $\tent=2/(\pi R)$ as in eq.~\eqref{spheretent} with $d=3$, and for the strip,
\beq
\label{eq:masslessd=3strip}
\delta \see =\frac{\delta E}{\tent}+\frac{\pi^{5/2}}{3\sqrt{2}\,\Gamma(3/4)^2}\,c^2 \,z_*\vol(\mathbb{R}),
\eeq
where $\tent=4\ell/(\pi^2z_*^2)$ with $z_*=\ell\Gamma(1/4)/2\sqrt{\pi}\Gamma(3/4)$, as in eq.~\eqref{striptent} with $d=3$. Via essentially the same arguments as those below eq.~\eqref{eq:masslessd=2}, for sufficiently small $c$ we can enter a regime where na\"ively we expect the FLEE of eq.~\eqref{flee2} to hold, but the two terms in eq.~\eqref{eq:masslessd=3sphere} or~\eqref{eq:masslessd=3strip} are of the same order. The FLEE of eq.~\eqref{flee2} is then clearly violated, as advertised.

\paragraph{Boundary Dimension $d=4$:}

The details of the holographic renormalisation for a massless scalar in asymptotically $AdS_5$ spacetimes appear in the appendix. In particular, plugging a Minkowski metric at the $AdS_5$ boundary and $\phi_0 = -ct$ into eqs.~\eqref{eq:d=4_relations} and~\eqref{eq:d=4SET} yields
\begin{subequations}
\begin{align}
	G_{xx}&=\frac{L^2}{z^2}\le[1+z^2\frac{2\pi}{3}\frac{G_N}{L^3}c^2\le(1-z^2\log\le(z^2/L^2\ri)2\pi\frac{G_N}{L^3}c^2\ri)+z^4 g_{xx}^{(4)}+\dots\ri],\\
	\le<T_{xx}\ri>&=\frac{L^3}{4\pi G_N} g_{xx}^{(4)}-\frac{5\pi}{18}\frac{G_N}{L^3}c^4,\label{eq:ads5holorenresultstxx} \\
	\le<T^\m{}_\m\ri>&=\frac{2\pi}{3}\frac{G_N}{L^3}c^4.\label{eq:ads5holorenresultstrace}
\end{align}
\end{subequations}
Plugging $\langle T_{xx}\rangle$ from eq.~\eqref{eq:ads5holorenresultstxx} into $\le<T^\m{}_\m\ri>$ from eq.~\eqref{eq:ads5holorenresultstrace} then gives $\langle T_{tt} \rangle  =3 \langle T_{xx} \rangle - \le<T^\m{}_\m\ri>$. As in the $d=2$ case above, a linearized perturbation $\delta G_{mn}$ is $\propto c^2$, so the change in the energy inside the sphere $\le|\vec{x}\ri|<R$ is
\begin{align}
	\d E=\frac{4}{3}\pi R^3\delta\le<T_{tt}\ri>=\frac{L^3}{G_N}R^3\delta g_{xx}^{(4)},
\end{align}
where $\delta g_{xx}^{(4)} \propto c^2$. As mentioned above, from eq.~\eqref{eq:perturbative_ee_2} we compute $\delta \see$ for the sphere,
\beq
\label{eq:masslessd=4}
\delta \see = \frac{\delta E}{\tent}-\frac{\pi^2}{9}\,c^2 \, R^2\le(5-6\log 2+6\log\le(\e/R\ri)\ri),
\eeq
with UV cut-off $z = \e$. The $\delta \see$ in eq.~\eqref{eq:masslessd=4} has some scheme dependence, via the term $\propto c^2 R^2 \log\le(\e/R\ri)$, such that re-scaling $\e$ shifts the terms $\propto c^2 R^2$. However, the \textit{coefficient} of the term $\propto c^2 R^2 \log\le(\e/R\ri)$ is invariant under re-scalings of $\e$, \textit{i.e.} is scheme-independent, and hence is physically meaningful.

Via essentially the same arguments as those below eq.~\eqref{eq:masslessd=2}, for sufficiently small $c$ we can enter a regime where na\"ively we expect the FLEE of eq.~\eqref{flee2} to hold, but all terms in eq.~\eqref{eq:masslessd=4} are of order $c^2 R^2$. The FLEE of eq.~\eqref{flee2} is then clearly violated, as advertised.

\subsection{Linear Spatial Dependence}

In this sub-section we consider the model of ref.~\cite{Andrade:2013gsa}, containing a $U(1)$ gauge field $A_m$ and massless scalars $\phi_I$ with $I=1,\ldots,d-1$ in asymptotically $AdS_{d+1}$ spacetime. We consider the solutions of ref.~\cite{Andrade:2013gsa} describing charged black branes with $\phi_I$ linear in a spatial direction $x$, dual to CFT states with non-zero chemical potential, $\mu$, and $x$-linear sources for a set of exactly marginal scalar operators $\mathcal{O}_I$. The main result of ref.~\cite{Andrade:2013gsa} was that the $x$-linear sources break translational symmetry in the CFT and hence produce the effects of momentum relaxation, such as a Drude peak in the $U(1)$ conductivity. We are instead interested in the $x$-linear sources as perturbations of the CFT at non-zero $\mu$. In $d=3$ we will show that the FLEE in eq.~\eqref{flee2} is violated, while both $\delta \see$ and $\delta E$ are independent of $t$ and $\vec{x}$, and hence will trivially obey a FLOER involving any CFT coordinate. Previous calculations of EE in the model of ref.~\cite{Andrade:2013gsa} appear for example in ref.~\cite{Mozaffara:2016iwm}.

The model of ref.~\cite{Andrade:2013gsa} has bulk action
\beq
S = \int d^{d+1}x \, \sqrt{-\textrm{det}(G_{mn})} \left[\frac{1}{16 \pi G_N}\left(R+\frac{d(d-1)}{L^2} - F^2\right) - \frac{1}{2} \sum_{I=1}^{d-1} (\partial \phi_I)^2 \right],
\eeq
where $F_{mn} = \partial_m A_n - \partial_n A_m$. We consider the solutions of ref.~\cite{Andrade:2013gsa} describing a static, charged black brane with scalar hair linear in $x$,
\begin{subequations}
\begin{align}
G_{mn} dx^m dx^n &= \frac{L^2}{u^2} \left(\frac{du^2}{f(u)} - f(u) dt^2 + d\vec{x}^2 \right),
\label{eq:andrade_withers_metric} \\
A_t &= \m \left[1 - \left(\frac{u}{u_h}\right)^{d-2} \right],
\\
\phi_I &= \vec{\a}_I \cdot \vec{x},\label{eq:masslessphisol}
\end{align}
\end{subequations}
with horizon at $u=u_h$ and all other components of $A_m$ vanishing. The constant vector $\vec{\a}_I$ in eq.~\eqref{eq:masslessphisol} has components $(\alpha_{I})_i$ with $i=1,\ldots,d-1$ defined such that
\beq
\sum_{I=1}^{d-1} (\a_{I})_i (\a_{I})_j = \a^2 \d_{ij},
\eeq
with constant \(\a^2\). The blackening function $f(u)$ appearing in the metric in eq.~\eqref{eq:andrade_withers_metric} is
\begin{subequations}
\label{eq:blackening}
\begin{align}
f(u) &= 1 - M \left(\frac{u}{L^2}\right)^d + \left(\frac{u_h \m}{\beta}\right)^2 \left(\frac{u}{u_h}\right)^{2(d-1)} - \frac{8 \pi G_N }{(d-2)}\,\a^2\,u^2,
\\
M &\equiv	\left[1 + \left(\frac{u_h\m}{\beta}\right)^2 - \frac{8 \pi G_N}{(d-2)}\, \a^2 \, u_h^2 \right] \left(\frac{L^2}{u_h}\right)^d,
\\
\beta^2 &\equiv \frac{d-1}{d-2} \frac{L^2}{2}.
\end{align}
\end{subequations}
When $\alpha^2=0$, this solution reduces to the $AdS_{d+1}$-Reissner-Nordstr\"om charged black brane.

We henceforth specialise to $d=3$, the case for which the holographic renormalisation of this model was performed in ref.~\cite{Andrade:2013gsa}. When $d=3$, the asymptotic change of coordinates
\beq
u = z - z^3\,2\pi G_N\a^2 - z^4\,\frac{M}{6L^6} + \co(z^5),
\eeq
brings the metric in eq.~\eqref{eq:andrade_withers_metric} into asymptotic FG form
\begin{subequations}
\label{eq:andrade_withers_metric_FG}
\begin{align}
G_{mn}dx^mdx^n &= \frac{L^2}{z^2} \left( dz^2 + g_{tt} dt^2 + g_{xx} d\vec{x}^2 \right),\\
g_{tt} & = -1+z^2 4 \pi G_N \a^2 + z^3\frac{2M}{3 L^6} + \co(z^4),\\
g_{xx} &=1+z^2 4 \pi G_N \a^2 + z^3\frac{M}{3 L^6} + \co(z^4).
\end{align}
\end{subequations}
The holographic renormalisation in ref.~\cite{Andrade:2013gsa} then gives for the energy density
\beq
\label{eq:andradewithersenergydensity}
\T{tt} = \frac{3L^2}{8 \p G_N} g_{xx}^{(3)} = \frac{M}{8 \pi G_N L^4}.
\eeq

For sufficiently small $\alpha^2$, we may treat the terms $\propto \alpha^2$ in eq.~\eqref{eq:blackening} as perturbations, and write $G_{mn} = G_{mn}^{(0)} + \delta G_{mn}$, with $G_{mn}^{(0)}$ the $AdS_4$-Reissner-Nordstr\"om metric, and $\delta G_{mn}$ of order $\alpha^2$. In particular,
\beq
\d g_{xx} = z^2\,4 \pi G_N \alpha^2 - z^3\,\frac{8 \pi G_N \alpha^2}{3 u_h} + \co(z^4).
\eeq
Using eq.~\eqref{eq:andradewithersenergydensity} we thus find the change in energy inside a spherical sub-region comes from the change in the order $z^3$ term in the FG asymptotics
\beq
\d E= (\pi R^2)\frac{3 L^2}{8\p G_N} \d g_{xx}^{(3)} = -  (\pi R^2)\frac{L^2 \a^2}{u_h},
\eeq
and the change in energy inside a strip sub-region is identical, but with $(\pi R^2) \to \ell \, \textrm{Vol}(\mathbb{R})$.

In contrast, $\delta \see$ depends on both $\delta g_{xx}^{(2)}$ and $\delta g_{xx}^{(3)}$. Indeed, applying the results of section~\ref{general}, we find for spherical and strip sub-regions, respectively,
\begin{subequations}
\label{eq:masslesslinearx}
	\begin{align}
	\d\see^\mathrm{sphere}
	&= \frac{\vol(S^1)}{8 G_N} \int_0^{z_\star} dz \left(\frac{L}{z}\right)^{2} r g_{xx}^{-1/2} \sqrt{1 + g_{xx} r'^2} \left(\frac{g_{xx} r'^2}{1 + g_{xx} r'^2} +1\right)
	\left(\d g^{(2)}_{xx} z^2 + \d g^{(3)}_{xx} z^3\ldots\right),
	\\
	\d\see^\mathrm{strip} 
	&= \frac{\vol(\mathbb{R}^1)}{4 G_N} \int_0^{z_\star} dz \left(\frac{L}{z}\right)^{2} g_{xx}^{-1/2} 
	\frac{(g_{xx}^\star z^2/g_{xx} z_{\star}^2)^{2} + 1}{\sqrt{1 - (g_{xx}^\star z^2/g_{xx} z_{\star}^2)^{2}}}
	\left( \d g^{(2)}_{xx} z^2 + \d g^{(3)}_{xx} z^3\ldots \right).
	\end{align}
\end{subequations}
In the $\delta \see$ in eq.~\eqref{eq:masslesslinearx}, a contribution $\propto \delta E$ can only possibly come from the terms involving $\delta g_{xx}^{(3)}$, so the terms involving $\delta g_{xx}^{(2)}$ represent violations of the FLEE of eq.~\eqref{flee2}. On the other hand, both $\delta E$ and $\delta \see$ are independent of $t$ and $\vec{x}$, so a FLOER involving any CFT coordinate is trivially obeyed, as advertised.

\section*{Acknowledgements}

We thank J.~Bhattacharya, A.~Karch, R.~Myers, and K.~Skenderis for useful conversations. A.~O'B. also thanks J.~Erdmenger and the University of W\"urzburg for hospitality during the completion of this work. A.~O'B. is a Royal Society University Research Fellow. The work of J.~P. was supported by a Clarendon scholarship from the Clarendon Fund and St John's College, Oxford, and by the European Research Council under the European Union's Seventh Framework Programme (ERC Grant agreement 307955). R.~R. acknowledges support from STFC through Consolidated Grant ST/L000296/1. The work of C.F.~U. was supported in part by the U.S.~Department of Energy under Grant No.~DE-SC0011637 and in part by the National Science Foundation under grant PHY-13-13986 and PHY-16-19926.

\appendix
\section{Holographic Renormalization of Massless Scalar}
\label{holoRG}

In this appendix we present results for the holographic renormalisation of a massless scalar field $\phi$ coupled to an asymptotically $AdS_4$ or $AdS_5$ metric $G_{mn}$. A massless scalar field $\phi$ and metric $G_{mn}$ are dual to a marginal scalar operator $\mathcal{O}$ and the stress-energy tensor $T_{\mu\nu}$, respectively. For a scalar of any mass coupled to gravity, a convenient form of Einstein's equations appear in ref.~\cite{deHaro:2000vlm}. For a massless scalar, we solve the Einstein's equations in ref.~\cite{deHaro:2000vlm} asymptotically\footnote{In our conventions the Riemann tensor has the opposite sign compared to that in ref.~\cite{deHaro:2000vlm}.}, and then compute $\langle \mathcal{O} \rangle$ and $\langle T_{\mu\nu} \rangle$ and the diffeomorphism and Weyl Ward identities in terms of the coefficients of $\phi$ and $G_{mn}$'s asymptotic expansions in eqs.~\eqref{eq:phiFGexp} and~\eqref{eq:metric_expansion}, respectively. We use the results for $\langle T_{\mu\nu}\rangle$ and $T^{\mu}_{~\mu}$ in sub-section~\ref{lineartime}, to compute the change in energy inside a CFT sub-region due to a $t$-linear source for $\mathcal{O}$.

In contrast to the body of the paper, in this appendix we choose units with $L \equiv 1$, and we use notation $\sqrt{-G} \equiv \sqrt{-\textrm{det} (G_{mn})}$, and similarly for other metrics.

\paragraph{Boundary Dimension $d=3$:} For a massless scalar field $\phi$ coupled to an asymptotically $AdS_4$ metric $G_{mn}$, we find
\begin{subequations} \label{eq:d=3_relations}
	\begin{align}
	\phi_2 &= \frac{1}{2} \nabla^2 \phi_0,
	\\
	g_{\mu\nu}^{(2)} &= - R_{\mu\nu}[g^{(0)}] + \frac{1}{4} R[g^{(0)}]
	+ 8 \pi G_N \partial_\mu \phi_0 \partial_\nu \phi_0 - 2 \pi G_N g_{\mu\nu}^{(0)}(\partial \phi_0)^2,
	\\
	\tr \, g^{(2)} &= - \frac{1}{4} R[g^{(0)}] + 2 \pi G_N (\partial \phi_0)^2,
	\\
	\tr \, g^{(3)} &= 0,
	\label{eq:d=3_tr_g3}
	\\
	\nabla^\nu g_{\mu\nu}^{(2)} &= \partial_\mu \tr g^{(2)} + 16 \pi G_N \phi_2 \partial_\mu \phi_0,
	\\
	\nabla^\nu g_{\mu\nu}^{(3)} &= 16 \pi G_N \phi_3 \partial_\mu \phi_0,
	\label{eq:d=3_div_g3}
	\end{align}
\end{subequations}
where $\nabla^\mu$ is with respect to \(g^{(0)}\), indices are raised and lowered with \(g^{(0)}\), and \(\tr g^{(N)} \equiv g^{(0)}_{\m\n} g^{(N)\m\n}\). The renormalised action is
\begin{align} \label{eq:d=3_s_ren}
S_\mathrm{ren} =  \lim_{\e\to 0} \Biggl\{&\frac{1}{16\pi G_N} \left[ \int d^4 x \, \sqrt{-G} \, (R + 6) 
+ 2 \int_{z=\e} d^3 x \, \sqrt{-\tilde{g}} \, K[\tilde{g}]\right]
\nonumber \\ &
- \frac{1}{2} \int d^4 x \, \sqrt{-G} \, G^{mn} \partial_m \phi \partial_n \phi
\nonumber \\ &
+ \frac{1}{16 \pi G_N} \int_{z=\e} d^3 x \, \sqrt{-\tilde{g}} \left(
4 + R[\tilde{g}] - 8 \pi G_N \tilde{g}^{\mu\nu} \partial_\mu \phi \partial_\nu \phi
\right)\Biggr\},
\end{align}
where $\tilde{g}_{\mu\nu}$ is the induced metric on a regulating cutoff surface $z = \e$, with extrinsic curvature $K[\tilde{g}]$ and Ricci scalar $R[\tilde{g}]$. The final line of eq.~\eqref{eq:d=3_s_ren} consists of counterterms at $z = \e$. Varying $S_{\textrm{ren}}$ in eq.~\eqref{eq:d=3_s_ren} with respect to sources, we obtain the one point functions
\begin{subequations}
	\begin{align}
	\langle \mathcal{O} \rangle &= 3 \, \phi_3,\\
	\langle T_{\mu\nu} \rangle &= \frac{3}{16\pi G_N} g_{\mu\nu}^{(3)},\label{eq:d=3SET}
	\end{align}
\end{subequations}
although the values of $\phi_3$ and $g_{\mu\nu}^{(3)}$ cannot be fixed by our near-boundary analysis alone. Eqns.~\eqref{eq:d=3_tr_g3} and~\eqref{eq:d=3_div_g3} yield the diffeomorphism and Weyl Ward identities, respectively,
\begin{subequations}
	\begin{align}
	\nabla^\m \langle T_{\m\n} \rangle &= \langle \co \rangle \partial_\n \f_0,
	\\
	\langle {T^\m}_{\m} \rangle &= 0.
	\end{align}
\end{subequations}

\paragraph{Boundary Dimension $d=4$:} For a massless scalar field $\phi$ coupled to an asymptotically $AdS_5$ metric $G_{mn}$, we find, with the same conventions as in eq.~\eqref{eq:d=3_relations},
\begin{subequations}
\label{eq:d=4_relations}
	\begin{align}
	\f_2 &= \frac{1}{4} \nabla^2 \f_0,
	\\
	\y_4 &= -\frac{1}{32} (\nabla^2)^2 \f_0 + \frac{1}{8} \frac{1}{\sqrt{-g^{(0)}}} \partial_\m (\sqrt{-g^{(0)}} \, g^{(2)\m\n} \partial_\n \f_0)
	\nonumber \\ &\phantom{=}
	 - \frac{1}{16} \partial_\m \tr \, g^{(2)} \, g^{(0)\m\n} \partial_\n \f_0- \frac{1}{16} \tr \, g^{(2)} \, \nabla^2 \f_0,
	\\
	g^{(2)}_{\m\n} &= - \frac{1}{2} R_{\m\n}[g^{(0)}] + \frac{1}{12} g^{(0)}_{\m\n} R[g^{(0)}] + 4 \p G_N \partial_\m \f_0 \partial_\n \f_0 - \frac{2 \p G_N}{3} g^{(0)}_{\m\n} (\partial \f_0)^2,
	\\
	h^{(4)}_{\m\n} &= + \frac{1}{4} R_{\m\n}[g^{(2)}] + \frac{1}{2} g^{(2)}_{\m\l} g_{~~~~\nu}^{(2)\lambda} - \frac{1}{8} g^{(0)}_{\m\n} \tr \, [(g^{(2)})^2] - \frac{1}{4}\p G_N g^{(0)}_{\m\n} (\nabla^2 \f_0)^2
	\nonumber \\ &\phantom{=}
	- \frac{1}{2}\p G_N (\partial_\m \f_0 \partial_\n \nabla^2 \f_0 + \partial_\m\nabla^2 \f_0 \partial_\n \f_0),
	\\
	\tr \, h^{(4)} &= 0,
	\\
	\nabla^\n h^{(4)}_{\m\n} &=  4 \p G_N \y_4 \partial_\m \nabla^2 \f_0,
	\\
	\tr \, g^{(4)} &= \frac{1}{4} \tr[(g^{(2)})^2]- \frac{1}{2}\p G_N (\nabla^2 \f_0)^2,
	\\
	\nabla^\n g^{(4)}_{\m\n} &= -\frac{1}{4} \partial_\m \tr [ (g^{(2)})^2] + 16 \p G_N \f_4 \partial_\m \f_0 - \frac{1}{2} \p G_N (\nabla^2 \f_0) \partial_\m (\nabla^2 \f_0),
	\end{align}
\end{subequations}
where $\psi_4$ is the coefficient of the $z^4 \log z^2$ term in $\phi$'s asymptotic expansion. The renormalised action is
\begin{align} \label{eq:d=4_s_ren}
S_\mathrm{ren} &= \lim_{\e\to 0} \Biggl\{\frac{1}{16\p G_N} \left[\int d^5 x \sqrt{-G}(R+12) + 2 \int_{z=\e} d^4 x \sqrt{-\tilde{g}} \, K[\tilde{g}] \right] \nonumber \\
&\phantom{=} - \frac{1}{2} \int d^5 x \sqrt{-G} \, G^{mn} \partial_m \f \partial_n \f
 \\ &\phantom{=}
+ \frac{1}{16\p G_N} \int_{z=\e} d^4 x \sqrt{-\tilde{g}} \left(
6 + \frac{1}{2} R[\tilde{g}] + 4\p G_N \tilde{g}^{\m\n}\partial_\m \f \partial_\n \f + a_{(4)} \e^2 \log \e
\right)\Biggr\}, \nonumber
\end{align}
where the final line consists of counterterms at $z =\e$, and
\begin{align} \label{eq:massless_a4}
a_{(4)} \equiv \frac{1}{\e^2} \biggl(
-\frac{1}{4} R^{\m\n}[\tilde{g}] R_{\m\n}[\tilde{g}] + \frac{1}{12} R[\tilde{g}]^2 + &4\p G_N R^{\m\n}[\tilde{g}] \partial_\m \f \partial_\n \f
\nonumber \\ &- \frac{136}{9} \p^2 G_N^2 (\tilde{g}^{\m\n} \partial_\m \partial_\n \f)^2 - \p G_N (\nabla^2_{\tilde{g}} \f)^2
\biggr),
\end{align}
where $R_{\m\n}[\tilde{g}]$ is the Ricci tensor of $\tilde{g}_{\mu\nu}$, and $\nabla^2_{\tilde{g}}$ is with respect to $\tilde{g}_{\mu\nu}$. Varying $S_{\textrm{ren}}$ in eq.~\eqref{eq:d=4_s_ren} with respect to sources, we obtain the one point functions
\begin{subequations}
	\begin{align}
	\langle \co \rangle &= 4 \f_4 + 6 \y_4 + \f_2 \tr g^{(2)}, \\
	\T{\m\n} &= \frac{1}{8\p G_N} \biggl(
	2 g^{(4)}_{\m\n} + 3 h^{(4)}_{\m\n} - g^{(2)}_{\m\l}  g^{(2)\lambda}_{~~~~\nu}
	+ \frac{1}{2} g^{(2)}_{\m\n} \tr \, g^{(2)} 
	\nonumber \\ &\phantom{= \frac{1}{8\p G_N} \biggl(}
	+ \frac{1}{2} g^{(0)}_{\m\n} \tr [ (g^{(2)})^2]- \frac{1}{4} g^{(0)}_{\m\n} [\tr g^{(2)}]^2 - g^{(0)}_{\m\n} \tr \, g^{(4)}
	\biggr).\label{eq:d=4SET}
	\end{align}
\end{subequations}
\bibliography{eefirst}

\begin{thebibliography}{64}%
\makeatletter
\providecommand \@ifxundefined [1]{%
 \@ifx{#1\undefined}
}%
\providecommand \@ifnum [1]{%
 \ifnum #1\expandafter \@firstoftwo
 \else \expandafter \@secondoftwo
 \fi
}%
\providecommand \@ifx [1]{%
 \ifx #1\expandafter \@firstoftwo
 \else \expandafter \@secondoftwo
 \fi
}%
\providecommand \natexlab [1]{#1}%
\providecommand \enquote  [1]{``#1''}%
\providecommand \bibnamefont  [1]{#1}%
\providecommand \bibfnamefont [1]{#1}%
\providecommand \citenamefont [1]{#1}%
\providecommand \href@noop [0]{\@secondoftwo}%
\providecommand \href [0]{\begingroup \@sanitize@url \@href}%
\providecommand \@href[1]{\@@startlink{#1}\@@href}%
\providecommand \@@href[1]{\endgroup#1\@@endlink}%
\providecommand \@sanitize@url [0]{\catcode `\\12\catcode `\$12\catcode
  `\&12\catcode `\#12\catcode `\^12\catcode `\_12\catcode `\%12\relax}%
\providecommand \@@startlink[1]{}%
\providecommand \@@endlink[0]{}%
\providecommand \url  [0]{\begingroup\@sanitize@url \@url }%
\providecommand \@url [1]{\endgroup\@href {#1}{\urlprefix }}%
\providecommand \urlprefix  [0]{URL }%
\providecommand \Eprint [0]{\href }%
\providecommand \doibase [0]{http://dx.doi.org/}%
\providecommand \selectlanguage [0]{\@gobble}%
\providecommand \bibinfo  [0]{\@secondoftwo}%
\providecommand \bibfield  [0]{\@secondoftwo}%
\providecommand \translation [1]{[#1]}%
\providecommand \BibitemOpen [0]{}%
\providecommand \bibitemStop [0]{}%
\providecommand \bibitemNoStop [0]{.\EOS\space}%
\providecommand \EOS [0]{\spacefactor3000\relax}%
\providecommand \BibitemShut  [1]{\csname bibitem#1\endcsname}%
\let\auto@bib@innerbib\@empty
\bibitem [{\citenamefont {Calabrese}\ and\ \citenamefont
  {Cardy}(2005)}]{Calabrese:2005in}%
  \BibitemOpen
  \bibfield  {author} {\bibinfo {author} {\bibfnamefont {Pasquale}\
  \bibnamefont {Calabrese}}\ and\ \bibinfo {author} {\bibfnamefont {John~L.}\
  \bibnamefont {Cardy}},\ }\bibfield  {title} {\enquote {\bibinfo {title}
  {{Evolution of entanglement entropy in one-dimensional systems}},}\ }\href
  {\doibase 10.1088/1742-5468/2005/04/P04010} {\bibfield  {journal} {\bibinfo
  {journal} {J. Stat. Mech.}\ }\textbf {\bibinfo {volume} {0504}},\ \bibinfo
  {pages} {P04010} (\bibinfo {year} {2005})},\ \Eprint
  {http://arxiv.org/abs/cond-mat/0503393} {arXiv:cond-mat/0503393 [cond-mat]}
  \BibitemShut {NoStop}%
\bibitem [{\citenamefont {Calabrese}\ and\ \citenamefont
  {Cardy}(2007)}]{Calabrese:2007mtj}%
  \BibitemOpen
  \bibfield  {author} {\bibinfo {author} {\bibfnamefont {Pasquale}\
  \bibnamefont {Calabrese}}\ and\ \bibinfo {author} {\bibfnamefont {John}\
  \bibnamefont {Cardy}},\ }\bibfield  {title} {\enquote {\bibinfo {title}
  {{Entanglement and correlation functions following a local quench: a
  conformal field theory approach}},}\ }\href {\doibase
  10.1088/1742-5468/2007/10/P10004} {\bibfield  {journal} {\bibinfo  {journal}
  {J. Stat. Mech.}\ }\textbf {\bibinfo {volume} {0710}},\ \bibinfo {pages}
  {P10004} (\bibinfo {year} {2007})},\ \Eprint {http://arxiv.org/abs/0708.3750}
  {arXiv:0708.3750 [quant-ph]} \BibitemShut {NoStop}%
\bibitem [{\citenamefont {Calabrese}\ and\ \citenamefont
  {Cardy}(2009)}]{Calabrese:2009qy}%
  \BibitemOpen
  \bibfield  {author} {\bibinfo {author} {\bibfnamefont {Pasquale}\
  \bibnamefont {Calabrese}}\ and\ \bibinfo {author} {\bibfnamefont {John}\
  \bibnamefont {Cardy}},\ }\bibfield  {title} {\enquote {\bibinfo {title}
  {{Entanglement entropy and conformal field theory}},}\ }\href {\doibase
  10.1088/1751-8113/42/50/504005} {\bibfield  {journal} {\bibinfo  {journal}
  {J. Phys.}\ }\textbf {\bibinfo {volume} {A42}},\ \bibinfo {pages} {504005}
  (\bibinfo {year} {2009})},\ \Eprint {http://arxiv.org/abs/0905.4013}
  {arXiv:0905.4013 [cond-mat.stat-mech]} \BibitemShut {NoStop}%
\bibitem [{\citenamefont {Calabrese}\ and\ \citenamefont
  {Cardy}(2016)}]{Calabrese:2016xau}%
  \BibitemOpen
  \bibfield  {author} {\bibinfo {author} {\bibfnamefont {Pasquale}\
  \bibnamefont {Calabrese}}\ and\ \bibinfo {author} {\bibfnamefont {John}\
  \bibnamefont {Cardy}},\ }\bibfield  {title} {\enquote {\bibinfo {title}
  {{Quantum quenches in 1+1 dimensional conformal field theories}},}\ }\href
  {\doibase 10.1088/1742-5468/2016/06/064003} {\bibfield  {journal} {\bibinfo
  {journal} {J. Stat. Mech.}\ }\textbf {\bibinfo {volume} {1606}},\ \bibinfo
  {pages} {064003} (\bibinfo {year} {2016})},\ \Eprint
  {http://arxiv.org/abs/1603.02889} {arXiv:1603.02889 [cond-mat.stat-mech]}
  \BibitemShut {NoStop}%
\bibitem [{\citenamefont {Liu}\ and\ \citenamefont
  {Suh}(2014{\natexlab{a}})}]{Liu:2013iza}%
  \BibitemOpen
  \bibfield  {author} {\bibinfo {author} {\bibfnamefont {Hong}\ \bibnamefont
  {Liu}}\ and\ \bibinfo {author} {\bibfnamefont {S.~Josephine}\ \bibnamefont
  {Suh}},\ }\bibfield  {title} {\enquote {\bibinfo {title} {{Entanglement
  Tsunami: Universal Scaling in Holographic Thermalization}},}\ }\href
  {\doibase 10.1103/PhysRevLett.112.011601} {\bibfield  {journal} {\bibinfo
  {journal} {Phys. Rev. Lett.}\ }\textbf {\bibinfo {volume} {112}},\ \bibinfo
  {pages} {011601} (\bibinfo {year} {2014}{\natexlab{a}})},\ \Eprint
  {http://arxiv.org/abs/1305.7244} {arXiv:1305.7244 [hep-th]} \BibitemShut
  {NoStop}%
\bibitem [{\citenamefont {Liu}\ and\ \citenamefont
  {Suh}(2014{\natexlab{b}})}]{Liu:2013qca}%
  \BibitemOpen
  \bibfield  {author} {\bibinfo {author} {\bibfnamefont {Hong}\ \bibnamefont
  {Liu}}\ and\ \bibinfo {author} {\bibfnamefont {S.~Josephine}\ \bibnamefont
  {Suh}},\ }\bibfield  {title} {\enquote {\bibinfo {title} {{Entanglement
  growth during thermalization in holographic systems}},}\ }\href {\doibase
  10.1103/PhysRevD.89.066012} {\bibfield  {journal} {\bibinfo  {journal} {Phys.
  Rev.}\ }\textbf {\bibinfo {volume} {D89}},\ \bibinfo {pages} {066012}
  (\bibinfo {year} {2014}{\natexlab{b}})},\ \Eprint
  {http://arxiv.org/abs/1311.1200} {arXiv:1311.1200 [hep-th]} \BibitemShut
  {NoStop}%
\bibitem [{\citenamefont {Casini}\ \emph
  {et~al.}(2016{\natexlab{a}})\citenamefont {Casini}, \citenamefont {Liu},\
  and\ \citenamefont {Mezei}}]{Casini:2015zua}%
  \BibitemOpen
  \bibfield  {author} {\bibinfo {author} {\bibfnamefont {Horacio}\ \bibnamefont
  {Casini}}, \bibinfo {author} {\bibfnamefont {Hong}\ \bibnamefont {Liu}}, \
  and\ \bibinfo {author} {\bibfnamefont {Márk}\ \bibnamefont {Mezei}},\
  }\bibfield  {title} {\enquote {\bibinfo {title} {{Spread of entanglement and
  causality}},}\ }\href {\doibase 10.1007/JHEP07(2016)077} {\bibfield
  {journal} {\bibinfo  {journal} {JHEP}\ }\textbf {\bibinfo {volume} {07}},\
  \bibinfo {pages} {077} (\bibinfo {year} {2016}{\natexlab{a}})},\ \Eprint
  {http://arxiv.org/abs/1509.05044} {arXiv:1509.05044 [hep-th]} \BibitemShut
  {NoStop}%
\bibitem [{\citenamefont {Avery}\ and\ \citenamefont
  {Paulos}(2014)}]{Avery:2014dba}%
  \BibitemOpen
  \bibfield  {author} {\bibinfo {author} {\bibfnamefont {Steven~G.}\
  \bibnamefont {Avery}}\ and\ \bibinfo {author} {\bibfnamefont {Miguel~F.}\
  \bibnamefont {Paulos}},\ }\bibfield  {title} {\enquote {\bibinfo {title}
  {{Universal Bounds on the Time Evolution of Entanglement Entropy}},}\ }\href
  {\doibase 10.1103/PhysRevLett.113.231604} {\bibfield  {journal} {\bibinfo
  {journal} {Phys. Rev. Lett.}\ }\textbf {\bibinfo {volume} {113}},\ \bibinfo
  {pages} {231604} (\bibinfo {year} {2014})},\ \Eprint
  {http://arxiv.org/abs/1407.0705} {arXiv:1407.0705 [hep-th]} \BibitemShut
  {NoStop}%
\bibitem [{\citenamefont {Hartman}\ and\ \citenamefont
  {Afkhami-Jeddi}(2015)}]{Hartman:2015apr}%
  \BibitemOpen
  \bibfield  {author} {\bibinfo {author} {\bibfnamefont {Thomas}\ \bibnamefont
  {Hartman}}\ and\ \bibinfo {author} {\bibfnamefont {Nima}\ \bibnamefont
  {Afkhami-Jeddi}},\ }\bibfield  {title} {\enquote {\bibinfo {title} {{Speed
  Limits for Entanglement}},}\ }\href@noop {} {\  (\bibinfo {year} {2015})},\
  \Eprint {http://arxiv.org/abs/1512.02695} {arXiv:1512.02695 [hep-th]}
  \BibitemShut {NoStop}%
\bibitem [{\citenamefont {Bhattacharya}\ \emph {et~al.}(2013)\citenamefont
  {Bhattacharya}, \citenamefont {Nozaki}, \citenamefont {Takayanagi},\ and\
  \citenamefont {Ugajin}}]{Bhattacharya:2012mi}%
  \BibitemOpen
  \bibfield  {author} {\bibinfo {author} {\bibfnamefont {Jyotirmoy}\
  \bibnamefont {Bhattacharya}}, \bibinfo {author} {\bibfnamefont {Masahiro}\
  \bibnamefont {Nozaki}}, \bibinfo {author} {\bibfnamefont {Tadashi}\
  \bibnamefont {Takayanagi}}, \ and\ \bibinfo {author} {\bibfnamefont
  {Tomonori}\ \bibnamefont {Ugajin}},\ }\bibfield  {title} {\enquote {\bibinfo
  {title} {{Thermodynamical Property of Entanglement Entropy for Excited
  States}},}\ }\href {\doibase 10.1103/PhysRevLett.110.091602} {\bibfield
  {journal} {\bibinfo  {journal} {Phys. Rev. Lett.}\ }\textbf {\bibinfo
  {volume} {110}},\ \bibinfo {pages} {091602} (\bibinfo {year} {2013})},\
  \Eprint {http://arxiv.org/abs/1212.1164} {arXiv:1212.1164} \BibitemShut
  {NoStop}%
\bibitem [{\citenamefont {Allahbakhshi}\ \emph {et~al.}(2013)\citenamefont
  {Allahbakhshi}, \citenamefont {Alishahiha},\ and\ \citenamefont
  {Naseh}}]{Allahbakhshi:2013rda}%
  \BibitemOpen
  \bibfield  {author} {\bibinfo {author} {\bibfnamefont {Davood}\ \bibnamefont
  {Allahbakhshi}}, \bibinfo {author} {\bibfnamefont {Mohsen}\ \bibnamefont
  {Alishahiha}}, \ and\ \bibinfo {author} {\bibfnamefont {Ali}\ \bibnamefont
  {Naseh}},\ }\bibfield  {title} {\enquote {\bibinfo {title} {{Entanglement
  Thermodynamics}},}\ }\href {\doibase 10.1007/JHEP08(2013)102} {\bibfield
  {journal} {\bibinfo  {journal} {JHEP}\ }\textbf {\bibinfo {volume} {08}},\
  \bibinfo {pages} {102} (\bibinfo {year} {2013})},\ \Eprint
  {http://arxiv.org/abs/1305.2728} {arXiv:1305.2728 [hep-th]} \BibitemShut
  {NoStop}%
\bibitem [{\citenamefont {Blanco}\ \emph {et~al.}(2013)\citenamefont {Blanco},
  \citenamefont {Casini}, \citenamefont {Hung},\ and\ \citenamefont
  {Myers}}]{Blanco:2013joa}%
  \BibitemOpen
  \bibfield  {author} {\bibinfo {author} {\bibfnamefont {David~D.}\
  \bibnamefont {Blanco}}, \bibinfo {author} {\bibfnamefont {Horacio}\
  \bibnamefont {Casini}}, \bibinfo {author} {\bibfnamefont {Ling-Yan}\
  \bibnamefont {Hung}}, \ and\ \bibinfo {author} {\bibfnamefont {Robert~C.}\
  \bibnamefont {Myers}},\ }\bibfield  {title} {\enquote {\bibinfo {title}
  {{Relative Entropy and Holography}},}\ }\href {\doibase
  10.1007/JHEP08(2013)060} {\bibfield  {journal} {\bibinfo  {journal} {JHEP}\
  }\textbf {\bibinfo {volume} {08}},\ \bibinfo {pages} {060} (\bibinfo {year}
  {2013})},\ \Eprint {http://arxiv.org/abs/1305.3182} {arXiv:1305.3182
  [hep-th]} \BibitemShut {NoStop}%
\bibitem [{\citenamefont {Wong}\ \emph {et~al.}(2013)\citenamefont {Wong},
  \citenamefont {Klich}, \citenamefont {Pando~Zayas},\ and\ \citenamefont
  {Vaman}}]{Wong:2013gua}%
  \BibitemOpen
  \bibfield  {author} {\bibinfo {author} {\bibfnamefont {Gabriel}\ \bibnamefont
  {Wong}}, \bibinfo {author} {\bibfnamefont {Israel}\ \bibnamefont {Klich}},
  \bibinfo {author} {\bibfnamefont {Leopoldo~A.}\ \bibnamefont {Pando~Zayas}},
  \ and\ \bibinfo {author} {\bibfnamefont {Diana}\ \bibnamefont {Vaman}},\
  }\bibfield  {title} {\enquote {\bibinfo {title} {{Entanglement Temperature
  and Entanglement Entropy of Excited States}},}\ }\href {\doibase
  10.1007/JHEP12(2013)020} {\bibfield  {journal} {\bibinfo  {journal} {JHEP}\
  }\textbf {\bibinfo {volume} {12}},\ \bibinfo {pages} {020} (\bibinfo {year}
  {2013})},\ \Eprint {http://arxiv.org/abs/1305.3291} {arXiv:1305.3291
  [hep-th]} \BibitemShut {NoStop}%
\bibitem [{\citenamefont {Casini}\ \emph {et~al.}(2011)\citenamefont {Casini},
  \citenamefont {Huerta},\ and\ \citenamefont {Myers}}]{Casini:2011kv}%
  \BibitemOpen
  \bibfield  {author} {\bibinfo {author} {\bibfnamefont {H.}~\bibnamefont
  {Casini}}, \bibinfo {author} {\bibfnamefont {M.}~\bibnamefont {Huerta}}, \
  and\ \bibinfo {author} {\bibfnamefont {R.}~\bibnamefont {Myers}},\ }\bibfield
   {title} {\enquote {\bibinfo {title} {{Towards a Derivation of Holographic
  Entanglement Entropy}},}\ }\href {\doibase 10.1007/JHEP05(2011)036}
  {\bibfield  {journal} {\bibinfo  {journal} {JHEP}\ }\textbf {\bibinfo
  {volume} {1105}},\ \bibinfo {pages} {036} (\bibinfo {year} {2011})},\ \Eprint
  {http://arxiv.org/abs/1102.0440} {arXiv:1102.0440 [hep-th]} \BibitemShut
  {NoStop}%
\bibitem [{\citenamefont {Ryu}\ and\ \citenamefont
  {Takayanagi}(2006{\natexlab{a}})}]{Ryu:2006bv}%
  \BibitemOpen
  \bibfield  {author} {\bibinfo {author} {\bibfnamefont {S.}~\bibnamefont
  {Ryu}}\ and\ \bibinfo {author} {\bibfnamefont {T.}~\bibnamefont
  {Takayanagi}},\ }\bibfield  {title} {\enquote {\bibinfo {title} {{Holographic
  Derivation of Entanglement Entropy from AdS/CFT}},}\ }\href {\doibase
  10.1103/PhysRevLett.96.181602} {\bibfield  {journal} {\bibinfo  {journal}
  {Phys.Rev.Lett.}\ }\textbf {\bibinfo {volume} {96}},\ \bibinfo {pages}
  {181602} (\bibinfo {year} {2006}{\natexlab{a}})},\ \Eprint
  {http://arxiv.org/abs/hep-th/0603001} {arXiv:hep-th/0603001 [hep-th]}
  \BibitemShut {NoStop}%
\bibitem [{\citenamefont {Ryu}\ and\ \citenamefont
  {Takayanagi}(2006{\natexlab{b}})}]{Ryu:2006ef}%
  \BibitemOpen
  \bibfield  {author} {\bibinfo {author} {\bibfnamefont {S.}~\bibnamefont
  {Ryu}}\ and\ \bibinfo {author} {\bibfnamefont {T.}~\bibnamefont
  {Takayanagi}},\ }\bibfield  {title} {\enquote {\bibinfo {title} {{Aspects of
  Holographic Entanglement Entropy}},}\ }\href {\doibase
  10.1088/1126-6708/2006/08/045} {\bibfield  {journal} {\bibinfo  {journal}
  {JHEP}\ }\textbf {\bibinfo {volume} {0608}},\ \bibinfo {pages} {045}
  (\bibinfo {year} {2006}{\natexlab{b}})},\ \Eprint
  {http://arxiv.org/abs/hep-th/0605073} {arXiv:hep-th/0605073 [hep-th]}
  \BibitemShut {NoStop}%
\bibitem [{\citenamefont {Hubeny}\ \emph {et~al.}(2007)\citenamefont {Hubeny},
  \citenamefont {Rangamani},\ and\ \citenamefont {Takayanagi}}]{Hubeny:2007xt}%
  \BibitemOpen
  \bibfield  {author} {\bibinfo {author} {\bibfnamefont {Veronika~E.}\
  \bibnamefont {Hubeny}}, \bibinfo {author} {\bibfnamefont {Mukund}\
  \bibnamefont {Rangamani}}, \ and\ \bibinfo {author} {\bibfnamefont {Tadashi}\
  \bibnamefont {Takayanagi}},\ }\bibfield  {title} {\enquote {\bibinfo {title}
  {{A Covariant holographic entanglement entropy proposal}},}\ }\href {\doibase
  10.1088/1126-6708/2007/07/062} {\bibfield  {journal} {\bibinfo  {journal}
  {JHEP}\ }\textbf {\bibinfo {volume} {07}},\ \bibinfo {pages} {062} (\bibinfo
  {year} {2007})},\ \Eprint {http://arxiv.org/abs/0705.0016} {arXiv:0705.0016
  [hep-th]} \BibitemShut {NoStop}%
\bibitem [{\citenamefont {Lewkowycz}\ and\ \citenamefont
  {Maldacena}(2013)}]{Lewkowycz:2013nqa}%
  \BibitemOpen
  \bibfield  {author} {\bibinfo {author} {\bibfnamefont {Aitor}\ \bibnamefont
  {Lewkowycz}}\ and\ \bibinfo {author} {\bibfnamefont {Juan}\ \bibnamefont
  {Maldacena}},\ }\bibfield  {title} {\enquote {\bibinfo {title} {{Generalized
  gravitational entropy}},}\ }\href {\doibase 10.1007/JHEP08(2013)090}
  {\bibfield  {journal} {\bibinfo  {journal} {JHEP}\ }\textbf {\bibinfo
  {volume} {08}},\ \bibinfo {pages} {090} (\bibinfo {year} {2013})},\ \Eprint
  {http://arxiv.org/abs/1304.4926} {arXiv:1304.4926 [hep-th]} \BibitemShut
  {NoStop}%
\bibitem [{\citenamefont {Dong}\ \emph
  {et~al.}(2016{\natexlab{a}})\citenamefont {Dong}, \citenamefont {Lewkowycz},\
  and\ \citenamefont {Rangamani}}]{Dong:2016hjy}%
  \BibitemOpen
  \bibfield  {author} {\bibinfo {author} {\bibfnamefont {Xi}~\bibnamefont
  {Dong}}, \bibinfo {author} {\bibfnamefont {Aitor}\ \bibnamefont {Lewkowycz}},
  \ and\ \bibinfo {author} {\bibfnamefont {Mukund}\ \bibnamefont {Rangamani}},\
  }\bibfield  {title} {\enquote {\bibinfo {title} {{Deriving covariant
  holographic entanglement}},}\ }\href {\doibase 10.1007/JHEP11(2016)028}
  {\bibfield  {journal} {\bibinfo  {journal} {JHEP}\ }\textbf {\bibinfo
  {volume} {11}},\ \bibinfo {pages} {028} (\bibinfo {year}
  {2016}{\natexlab{a}})},\ \Eprint {http://arxiv.org/abs/1607.07506}
  {arXiv:1607.07506 [hep-th]} \BibitemShut {NoStop}%
\bibitem [{\citenamefont {Chamblin}\ \emph {et~al.}(1999)\citenamefont
  {Chamblin}, \citenamefont {Emparan}, \citenamefont {Johnson},\ and\
  \citenamefont {Myers}}]{Chamblin:1999tk}%
  \BibitemOpen
  \bibfield  {author} {\bibinfo {author} {\bibfnamefont {A.}~\bibnamefont
  {Chamblin}}, \bibinfo {author} {\bibfnamefont {R.}~\bibnamefont {Emparan}},
  \bibinfo {author} {\bibfnamefont {C.V.}\ \bibnamefont {Johnson}}, \ and\
  \bibinfo {author} {\bibfnamefont {R.C.}\ \bibnamefont {Myers}},\ }\bibfield
  {title} {\enquote {\bibinfo {title} {{Charged AdS Black Holes and
  Catastrophic Holography}},}\ }\href {\doibase 10.1103/PhysRevD.60.064018}
  {\bibfield  {journal} {\bibinfo  {journal} {Phys.Rev.}\ }\textbf {\bibinfo
  {volume} {D60}},\ \bibinfo {pages} {064018} (\bibinfo {year} {1999})},\
  \Eprint {http://arxiv.org/abs/hep-th/9902170} {arXiv:hep-th/9902170 [hep-th]}
  \BibitemShut {NoStop}%
\bibitem [{\citenamefont {Cvetic}\ \emph {et~al.}(1999)\citenamefont {Cvetic},
  \citenamefont {Duff}, \citenamefont {Hoxha}, \citenamefont {Liu},
  \citenamefont {Lu}, \citenamefont {Lu}, \citenamefont {Martinez-Acosta},
  \citenamefont {Pope}, \citenamefont {Sati},\ and\ \citenamefont
  {Tran}}]{Cvetic:1999xp}%
  \BibitemOpen
  \bibfield  {author} {\bibinfo {author} {\bibfnamefont {Mirjam}\ \bibnamefont
  {Cvetic}}, \bibinfo {author} {\bibfnamefont {M.~J.}\ \bibnamefont {Duff}},
  \bibinfo {author} {\bibfnamefont {P.}~\bibnamefont {Hoxha}}, \bibinfo
  {author} {\bibfnamefont {James~T.}\ \bibnamefont {Liu}}, \bibinfo {author}
  {\bibfnamefont {Hong}\ \bibnamefont {Lu}}, \bibinfo {author} {\bibfnamefont
  {J.~X.}\ \bibnamefont {Lu}}, \bibinfo {author} {\bibfnamefont
  {R.}~\bibnamefont {Martinez-Acosta}}, \bibinfo {author} {\bibfnamefont
  {C.~N.}\ \bibnamefont {Pope}}, \bibinfo {author} {\bibfnamefont
  {H.}~\bibnamefont {Sati}}, \ and\ \bibinfo {author} {\bibfnamefont {Tuan~A.}\
  \bibnamefont {Tran}},\ }\bibfield  {title} {\enquote {\bibinfo {title}
  {{Embedding AdS black holes in ten-dimensions and eleven-dimensions}},}\
  }\href {\doibase 10.1016/S0550-3213(99)00419-8} {\bibfield  {journal}
  {\bibinfo  {journal} {Nucl. Phys.}\ }\textbf {\bibinfo {volume} {B558}},\
  \bibinfo {pages} {96--126} (\bibinfo {year} {1999})},\ \Eprint
  {http://arxiv.org/abs/hep-th/9903214} {arXiv:hep-th/9903214 [hep-th]}
  \BibitemShut {NoStop}%
\bibitem [{\citenamefont {Aharony}\ \emph {et~al.}(2008)\citenamefont
  {Aharony}, \citenamefont {Bergman}, \citenamefont {Jafferis},\ and\
  \citenamefont {Maldacena}}]{Aharony:2008ug}%
  \BibitemOpen
  \bibfield  {author} {\bibinfo {author} {\bibfnamefont {O.}~\bibnamefont
  {Aharony}}, \bibinfo {author} {\bibfnamefont {O.}~\bibnamefont {Bergman}},
  \bibinfo {author} {\bibfnamefont {D.}~\bibnamefont {Jafferis}}, \ and\
  \bibinfo {author} {\bibfnamefont {J.}~\bibnamefont {Maldacena}},\ }\bibfield
  {title} {\enquote {\bibinfo {title} {{N=6 Superconformal Chern-Simons-matter
  Theories, M2-branes and Their Gravity Duals}},}\ }\href {\doibase
  10.1088/1126-6708/2008/10/091} {\bibfield  {journal} {\bibinfo  {journal}
  {JHEP}\ }\textbf {\bibinfo {volume} {10}},\ \bibinfo {pages} {091} (\bibinfo
  {year} {2008})},\ \Eprint {http://arxiv.org/abs/0806.1218} {arXiv:0806.1218
  [hep-th]} \BibitemShut {NoStop}%
\bibitem [{\citenamefont {Horowitz}\ \emph {et~al.}(2013)\citenamefont
  {Horowitz}, \citenamefont {Iqbal},\ and\ \citenamefont
  {Santos}}]{Horowitz:2013mia}%
  \BibitemOpen
  \bibfield  {author} {\bibinfo {author} {\bibfnamefont {Gary~T.}\ \bibnamefont
  {Horowitz}}, \bibinfo {author} {\bibfnamefont {Nabil}\ \bibnamefont {Iqbal}},
  \ and\ \bibinfo {author} {\bibfnamefont {Jorge~E.}\ \bibnamefont {Santos}},\
  }\bibfield  {title} {\enquote {\bibinfo {title} {{Simple holographic model of
  nonlinear conductivity}},}\ }\href {\doibase 10.1103/PhysRevD.88.126002}
  {\bibfield  {journal} {\bibinfo  {journal} {Phys. Rev.}\ }\textbf {\bibinfo
  {volume} {D88}},\ \bibinfo {pages} {126002} (\bibinfo {year} {2013})},\
  \Eprint {http://arxiv.org/abs/1309.5088} {arXiv:1309.5088 [hep-th]}
  \BibitemShut {NoStop}%
\bibitem [{\citenamefont {de~Haro}\ \emph
  {et~al.}(2001{\natexlab{a}})\citenamefont {de~Haro}, \citenamefont
  {Solodukhin},\ and\ \citenamefont {Skenderis}}]{deHaro:2000vlm}%
  \BibitemOpen
  \bibfield  {author} {\bibinfo {author} {\bibfnamefont {Sebastian}\
  \bibnamefont {de~Haro}}, \bibinfo {author} {\bibfnamefont {Sergey~N.}\
  \bibnamefont {Solodukhin}}, \ and\ \bibinfo {author} {\bibfnamefont {Kostas}\
  \bibnamefont {Skenderis}},\ }\bibfield  {title} {\enquote {\bibinfo {title}
  {{Holographic reconstruction of space-time and renormalization in the AdS /
  CFT correspondence}},}\ }\href {\doibase 10.1007/s002200100381} {\bibfield
  {journal} {\bibinfo  {journal} {Commun. Math. Phys.}\ }\textbf {\bibinfo
  {volume} {217}},\ \bibinfo {pages} {595--622} (\bibinfo {year}
  {2001}{\natexlab{a}})},\ \Eprint {http://arxiv.org/abs/hep-th/0002230}
  {arXiv:hep-th/0002230 [hep-th]} \BibitemShut {NoStop}%
\bibitem [{\citenamefont {Andrade}\ and\ \citenamefont
  {Withers}(2014)}]{Andrade:2013gsa}%
  \BibitemOpen
  \bibfield  {author} {\bibinfo {author} {\bibfnamefont {Tomas}\ \bibnamefont
  {Andrade}}\ and\ \bibinfo {author} {\bibfnamefont {Benjamin}\ \bibnamefont
  {Withers}},\ }\bibfield  {title} {\enquote {\bibinfo {title} {{A simple
  holographic model of momentum relaxation}},}\ }\href {\doibase
  10.1007/JHEP05(2014)101} {\bibfield  {journal} {\bibinfo  {journal} {JHEP}\
  }\textbf {\bibinfo {volume} {05}},\ \bibinfo {pages} {101} (\bibinfo {year}
  {2014})},\ \Eprint {http://arxiv.org/abs/1311.5157} {arXiv:1311.5157
  [hep-th]} \BibitemShut {NoStop}%
\bibitem [{\citenamefont {Karch}\ \emph {et~al.}(2009)\citenamefont {Karch},
  \citenamefont {O'Bannon},\ and\ \citenamefont {Thompson}}]{Karch:2008uy}%
  \BibitemOpen
  \bibfield  {author} {\bibinfo {author} {\bibfnamefont {Andreas}\ \bibnamefont
  {Karch}}, \bibinfo {author} {\bibfnamefont {Andy}\ \bibnamefont {O'Bannon}},
  \ and\ \bibinfo {author} {\bibfnamefont {Ethan}\ \bibnamefont {Thompson}},\
  }\bibfield  {title} {\enquote {\bibinfo {title} {{The Stress-Energy Tensor of
  Flavor Fields from AdS/CFT}},}\ }\href {\doibase
  10.1088/1126-6708/2009/04/021} {\bibfield  {journal} {\bibinfo  {journal}
  {JHEP}\ }\textbf {\bibinfo {volume} {04}},\ \bibinfo {pages} {021} (\bibinfo
  {year} {2009})},\ \Eprint {http://arxiv.org/abs/0812.3629} {arXiv:0812.3629
  [hep-th]} \BibitemShut {NoStop}%
\bibitem [{\citenamefont {Rangamani}\ \emph {et~al.}(2015)\citenamefont
  {Rangamani}, \citenamefont {Rozali},\ and\ \citenamefont
  {Wong}}]{Rangamani:2015sha}%
  \BibitemOpen
  \bibfield  {author} {\bibinfo {author} {\bibfnamefont {Mukund}\ \bibnamefont
  {Rangamani}}, \bibinfo {author} {\bibfnamefont {Moshe}\ \bibnamefont
  {Rozali}}, \ and\ \bibinfo {author} {\bibfnamefont {Anson}\ \bibnamefont
  {Wong}},\ }\bibfield  {title} {\enquote {\bibinfo {title} {{Driven
  Holographic CFTs}},}\ }\href {\doibase 10.1007/JHEP04(2015)093} {\bibfield
  {journal} {\bibinfo  {journal} {JHEP}\ }\textbf {\bibinfo {volume} {04}},\
  \bibinfo {pages} {093} (\bibinfo {year} {2015})},\ \Eprint
  {http://arxiv.org/abs/1502.05726} {arXiv:1502.05726 [hep-th]} \BibitemShut
  {NoStop}%
\bibitem [{\citenamefont {Casini}\ \emph
  {et~al.}(2016{\natexlab{b}})\citenamefont {Casini}, \citenamefont {Teste},\
  and\ \citenamefont {Torroba}}]{Casini:2016udt}%
  \BibitemOpen
  \bibfield  {author} {\bibinfo {author} {\bibfnamefont {Horacio}\ \bibnamefont
  {Casini}}, \bibinfo {author} {\bibfnamefont {Eduardo}\ \bibnamefont {Teste}},
  \ and\ \bibinfo {author} {\bibfnamefont {Gonzalo}\ \bibnamefont {Torroba}},\
  }\bibfield  {title} {\enquote {\bibinfo {title} {{Relative entropy and the RG
  flow}},}\ }\href@noop {} {\  (\bibinfo {year} {2016}{\natexlab{b}})},\
  \Eprint {http://arxiv.org/abs/1611.00016} {arXiv:1611.00016 [hep-th]}
  \BibitemShut {NoStop}%
\bibitem [{\citenamefont {Carracedo}\ \emph {et~al.}(2016)\citenamefont
  {Carracedo}, \citenamefont {Mas}, \citenamefont {Musso},\ and\ \citenamefont
  {Serantes}}]{Carracedo:2016qrf}%
  \BibitemOpen
  \bibfield  {author} {\bibinfo {author} {\bibfnamefont {Pablo}\ \bibnamefont
  {Carracedo}}, \bibinfo {author} {\bibfnamefont {Javier}\ \bibnamefont {Mas}},
  \bibinfo {author} {\bibfnamefont {Daniele}\ \bibnamefont {Musso}}, \ and\
  \bibinfo {author} {\bibfnamefont {Alexandre}\ \bibnamefont {Serantes}},\
  }\bibfield  {title} {\enquote {\bibinfo {title} {{Adiabatic pumping solutions
  in global AdS}},}\ }\href@noop {} {\  (\bibinfo {year} {2016})},\ \Eprint
  {http://arxiv.org/abs/1612.07701} {arXiv:1612.07701 [hep-th]} \BibitemShut
  {NoStop}%
\bibitem [{\citenamefont {Czech}\ \emph {et~al.}(2012)\citenamefont {Czech},
  \citenamefont {Karczmarek}, \citenamefont {Nogueira},\ and\ \citenamefont
  {Van~Raamsdonk}}]{Czech:2012bh}%
  \BibitemOpen
  \bibfield  {author} {\bibinfo {author} {\bibfnamefont {Bartlomiej}\
  \bibnamefont {Czech}}, \bibinfo {author} {\bibfnamefont {Joanna~L.}\
  \bibnamefont {Karczmarek}}, \bibinfo {author} {\bibfnamefont {Fernando}\
  \bibnamefont {Nogueira}}, \ and\ \bibinfo {author} {\bibfnamefont {Mark}\
  \bibnamefont {Van~Raamsdonk}},\ }\bibfield  {title} {\enquote {\bibinfo
  {title} {{The Gravity Dual of a Density Matrix}},}\ }\href {\doibase
  10.1088/0264-9381/29/15/155009} {\bibfield  {journal} {\bibinfo  {journal}
  {Class. Quant. Grav.}\ }\textbf {\bibinfo {volume} {29}},\ \bibinfo {pages}
  {155009} (\bibinfo {year} {2012})},\ \Eprint {http://arxiv.org/abs/1204.1330}
  {arXiv:1204.1330 [hep-th]} \BibitemShut {NoStop}%
\bibitem [{\citenamefont {Headrick}\ \emph {et~al.}(2014)\citenamefont
  {Headrick}, \citenamefont {Hubeny}, \citenamefont {Lawrence},\ and\
  \citenamefont {Rangamani}}]{Headrick:2014cta}%
  \BibitemOpen
  \bibfield  {author} {\bibinfo {author} {\bibfnamefont {Matthew}\ \bibnamefont
  {Headrick}}, \bibinfo {author} {\bibfnamefont {Veronika~E.}\ \bibnamefont
  {Hubeny}}, \bibinfo {author} {\bibfnamefont {Albion}\ \bibnamefont
  {Lawrence}}, \ and\ \bibinfo {author} {\bibfnamefont {Mukund}\ \bibnamefont
  {Rangamani}},\ }\bibfield  {title} {\enquote {\bibinfo {title} {{Causality \&
  holographic entanglement entropy}},}\ }\href {\doibase
  10.1007/JHEP12(2014)162} {\bibfield  {journal} {\bibinfo  {journal} {JHEP}\
  }\textbf {\bibinfo {volume} {12}},\ \bibinfo {pages} {162} (\bibinfo {year}
  {2014})},\ \Eprint {http://arxiv.org/abs/1408.6300} {arXiv:1408.6300
  [hep-th]} \BibitemShut {NoStop}%
\bibitem [{\citenamefont {Jafferis}\ and\ \citenamefont
  {Suh}(2016)}]{Jafferis:2014lza}%
  \BibitemOpen
  \bibfield  {author} {\bibinfo {author} {\bibfnamefont {Daniel~L.}\
  \bibnamefont {Jafferis}}\ and\ \bibinfo {author} {\bibfnamefont
  {S.~Josephine}\ \bibnamefont {Suh}},\ }\bibfield  {title} {\enquote {\bibinfo
  {title} {{The Gravity Duals of Modular Hamiltonians}},}\ }\href {\doibase
  10.1007/JHEP09(2016)068} {\bibfield  {journal} {\bibinfo  {journal} {JHEP}\
  }\textbf {\bibinfo {volume} {09}},\ \bibinfo {pages} {068} (\bibinfo {year}
  {2016})},\ \Eprint {http://arxiv.org/abs/1412.8465} {arXiv:1412.8465
  [hep-th]} \BibitemShut {NoStop}%
\bibitem [{\citenamefont {Jafferis}\ \emph {et~al.}(2016)\citenamefont
  {Jafferis}, \citenamefont {Lewkowycz}, \citenamefont {Maldacena},\ and\
  \citenamefont {Suh}}]{Jafferis:2015del}%
  \BibitemOpen
  \bibfield  {author} {\bibinfo {author} {\bibfnamefont {Daniel~L.}\
  \bibnamefont {Jafferis}}, \bibinfo {author} {\bibfnamefont {Aitor}\
  \bibnamefont {Lewkowycz}}, \bibinfo {author} {\bibfnamefont {Juan}\
  \bibnamefont {Maldacena}}, \ and\ \bibinfo {author} {\bibfnamefont
  {S.~Josephine}\ \bibnamefont {Suh}},\ }\bibfield  {title} {\enquote {\bibinfo
  {title} {{Relative entropy equals bulk relative entropy}},}\ }\href {\doibase
  10.1007/JHEP06(2016)004} {\bibfield  {journal} {\bibinfo  {journal} {JHEP}\
  }\textbf {\bibinfo {volume} {06}},\ \bibinfo {pages} {004} (\bibinfo {year}
  {2016})},\ \Eprint {http://arxiv.org/abs/1512.06431} {arXiv:1512.06431
  [hep-th]} \BibitemShut {NoStop}%
\bibitem [{\citenamefont {Dong}\ \emph
  {et~al.}(2016{\natexlab{b}})\citenamefont {Dong}, \citenamefont {Harlow},\
  and\ \citenamefont {Wall}}]{Dong:2016eik}%
  \BibitemOpen
  \bibfield  {author} {\bibinfo {author} {\bibfnamefont {Xi}~\bibnamefont
  {Dong}}, \bibinfo {author} {\bibfnamefont {Daniel}\ \bibnamefont {Harlow}}, \
  and\ \bibinfo {author} {\bibfnamefont {Aron~C.}\ \bibnamefont {Wall}},\
  }\bibfield  {title} {\enquote {\bibinfo {title} {{Reconstruction of Bulk
  Operators within the Entanglement Wedge in Gauge-Gravity Duality}},}\ }\href
  {\doibase 10.1103/PhysRevLett.117.021601} {\bibfield  {journal} {\bibinfo
  {journal} {Phys. Rev. Lett.}\ }\textbf {\bibinfo {volume} {117}},\ \bibinfo
  {pages} {021601} (\bibinfo {year} {2016}{\natexlab{b}})},\ \Eprint
  {http://arxiv.org/abs/1601.05416} {arXiv:1601.05416 [hep-th]} \BibitemShut
  {NoStop}%
\bibitem [{\citenamefont {Lokhande}\ \emph {et~al.}(2017)\citenamefont
  {Lokhande}, \citenamefont {Oling},\ and\ \citenamefont
  {Pedraza}}]{Lokhande:2017jik}%
  \BibitemOpen
  \bibfield  {author} {\bibinfo {author} {\bibfnamefont {Sagar~F.}\
  \bibnamefont {Lokhande}}, \bibinfo {author} {\bibfnamefont {Gerben W.~J.}\
  \bibnamefont {Oling}}, \ and\ \bibinfo {author} {\bibfnamefont {Juan~F.}\
  \bibnamefont {Pedraza}},\ }\bibfield  {title} {\enquote {\bibinfo {title}
  {{Linear response of entanglement entropy from holography}},}\ }\href@noop {}
  {\  (\bibinfo {year} {2017})},\ \Eprint {http://arxiv.org/abs/1705.10324}
  {arXiv:1705.10324 [hep-th]} \BibitemShut {NoStop}%
\bibitem [{\citenamefont {Nozaki}\ \emph {et~al.}(2013)\citenamefont {Nozaki},
  \citenamefont {Numasawa}, \citenamefont {Prudenziati},\ and\ \citenamefont
  {Takayanagi}}]{Nozaki:2013vta}%
  \BibitemOpen
  \bibfield  {author} {\bibinfo {author} {\bibfnamefont {Masahiro}\
  \bibnamefont {Nozaki}}, \bibinfo {author} {\bibfnamefont {Tokiro}\
  \bibnamefont {Numasawa}}, \bibinfo {author} {\bibfnamefont {Andrea}\
  \bibnamefont {Prudenziati}}, \ and\ \bibinfo {author} {\bibfnamefont
  {Tadashi}\ \bibnamefont {Takayanagi}},\ }\bibfield  {title} {\enquote
  {\bibinfo {title} {{Dynamics of Entanglement Entropy from Einstein
  Equation}},}\ }\href {\doibase 10.1103/PhysRevD.88.026012} {\bibfield
  {journal} {\bibinfo  {journal} {Phys. Rev.}\ }\textbf {\bibinfo {volume}
  {D88}},\ \bibinfo {pages} {026012} (\bibinfo {year} {2013})},\ \Eprint
  {http://arxiv.org/abs/1304.7100} {arXiv:1304.7100 [hep-th]} \BibitemShut
  {NoStop}%
\bibitem [{\citenamefont {Chang}\ and\ \citenamefont
  {Karch}(2014)}]{Chang:2013mca}%
  \BibitemOpen
  \bibfield  {author} {\bibinfo {author} {\bibfnamefont {Han-Chih}\
  \bibnamefont {Chang}}\ and\ \bibinfo {author} {\bibfnamefont {Andreas}\
  \bibnamefont {Karch}},\ }\bibfield  {title} {\enquote {\bibinfo {title}
  {{Entanglement Entropy for Probe Branes}},}\ }\href {\doibase
  10.1007/JHEP01(2014)180} {\bibfield  {journal} {\bibinfo  {journal} {JHEP}\
  }\textbf {\bibinfo {volume} {01}},\ \bibinfo {pages} {180} (\bibinfo {year}
  {2014})},\ \Eprint {http://arxiv.org/abs/1307.5325} {arXiv:1307.5325
  [hep-th]} \BibitemShut {NoStop}%
\bibitem [{\citenamefont {Lashkari}\ \emph {et~al.}(2014)\citenamefont
  {Lashkari}, \citenamefont {McDermott},\ and\ \citenamefont
  {Van~Raamsdonk}}]{Lashkari:2013koa}%
  \BibitemOpen
  \bibfield  {author} {\bibinfo {author} {\bibfnamefont {Nima}\ \bibnamefont
  {Lashkari}}, \bibinfo {author} {\bibfnamefont {Michael~B.}\ \bibnamefont
  {McDermott}}, \ and\ \bibinfo {author} {\bibfnamefont {Mark}\ \bibnamefont
  {Van~Raamsdonk}},\ }\bibfield  {title} {\enquote {\bibinfo {title}
  {{Gravitational dynamics from entanglement 'thermodynamics'}},}\ }\href
  {\doibase 10.1007/JHEP04(2014)195} {\bibfield  {journal} {\bibinfo  {journal}
  {JHEP}\ }\textbf {\bibinfo {volume} {04}},\ \bibinfo {pages} {195} (\bibinfo
  {year} {2014})},\ \Eprint {http://arxiv.org/abs/1308.3716} {arXiv:1308.3716
  [hep-th]} \BibitemShut {NoStop}%
\bibitem [{\citenamefont {Karch}\ and\ \citenamefont
  {Katz}(2002)}]{Karch:2002sh}%
  \BibitemOpen
  \bibfield  {author} {\bibinfo {author} {\bibfnamefont {Andreas}\ \bibnamefont
  {Karch}}\ and\ \bibinfo {author} {\bibfnamefont {Emanuel}\ \bibnamefont
  {Katz}},\ }\bibfield  {title} {\enquote {\bibinfo {title} {{Adding flavor to
  AdS/CFT}},}\ }\href@noop {} {\bibfield  {journal} {\bibinfo  {journal}
  {JHEP}\ }\textbf {\bibinfo {volume} {06}},\ \bibinfo {pages} {043} (\bibinfo
  {year} {2002})},\ \Eprint {http://arxiv.org/abs/hep-th/0205236}
  {arXiv:hep-th/0205236} \BibitemShut {NoStop}%
\bibitem [{\citenamefont {Maldacena}(1998)}]{Maldacena:1997re}%
  \BibitemOpen
  \bibfield  {author} {\bibinfo {author} {\bibfnamefont {Juan~Martin}\
  \bibnamefont {Maldacena}},\ }\bibfield  {title} {\enquote {\bibinfo {title}
  {{The Large N Limit of Superconformal Field Theories and Supergravity}},}\
  }\href@noop {} {\bibfield  {journal} {\bibinfo  {journal} {Adv. Theor. Math.
  Phys.}\ }\textbf {\bibinfo {volume} {2}},\ \bibinfo {pages} {231--252}
  (\bibinfo {year} {1998})},\ \Eprint {http://arxiv.org/abs/hep-th/9711200}
  {arXiv:hep-th/9711200} \BibitemShut {NoStop}%
\bibitem [{\citenamefont {Kontoudi}\ and\ \citenamefont
  {Policastro}(2014)}]{Kontoudi:2013rla}%
  \BibitemOpen
  \bibfield  {author} {\bibinfo {author} {\bibfnamefont {Konstantina}\
  \bibnamefont {Kontoudi}}\ and\ \bibinfo {author} {\bibfnamefont {Giuseppe}\
  \bibnamefont {Policastro}},\ }\bibfield  {title} {\enquote {\bibinfo {title}
  {{Flavor corrections to the entanglement entropy}},}\ }\href {\doibase
  10.1007/JHEP01(2014)043} {\bibfield  {journal} {\bibinfo  {journal} {JHEP}\
  }\textbf {\bibinfo {volume} {01}},\ \bibinfo {pages} {043} (\bibinfo {year}
  {2014})},\ \Eprint {http://arxiv.org/abs/1310.4549} {arXiv:1310.4549
  [hep-th]} \BibitemShut {NoStop}%
\bibitem [{\citenamefont {Karch}\ and\ \citenamefont
  {Uhlemann}(2014)}]{Karch:2014ufa}%
  \BibitemOpen
  \bibfield  {author} {\bibinfo {author} {\bibfnamefont {Andreas}\ \bibnamefont
  {Karch}}\ and\ \bibinfo {author} {\bibfnamefont {Christoph~F.}\ \bibnamefont
  {Uhlemann}},\ }\bibfield  {title} {\enquote {\bibinfo {title} {{Generalized
  gravitational entropy of probe branes: flavor entanglement
  holographically}},}\ }\href {\doibase 10.1007/JHEP05(2014)017} {\bibfield
  {journal} {\bibinfo  {journal} {JHEP}\ }\textbf {\bibinfo {volume} {05}},\
  \bibinfo {pages} {017} (\bibinfo {year} {2014})},\ \Eprint
  {http://arxiv.org/abs/1402.4497} {arXiv:1402.4497 [hep-th]} \BibitemShut
  {NoStop}%
\bibitem [{\citenamefont {Hashimoto}\ and\ \citenamefont
  {Oka}(2013)}]{Hashimoto:2013mua}%
  \BibitemOpen
  \bibfield  {author} {\bibinfo {author} {\bibfnamefont {Koji}\ \bibnamefont
  {Hashimoto}}\ and\ \bibinfo {author} {\bibfnamefont {Takashi}\ \bibnamefont
  {Oka}},\ }\bibfield  {title} {\enquote {\bibinfo {title} {{Vacuum Instability
  in Electric Fields via AdS/CFT: Euler-Heisenberg Lagrangian and Planckian
  Thermalization}},}\ }\href {\doibase 10.1007/JHEP10(2013)116} {\bibfield
  {journal} {\bibinfo  {journal} {JHEP}\ }\textbf {\bibinfo {volume} {10}},\
  \bibinfo {pages} {116} (\bibinfo {year} {2013})},\ \Eprint
  {http://arxiv.org/abs/1307.7423} {arXiv:1307.7423 [hep-th]} \BibitemShut
  {NoStop}%
\bibitem [{\citenamefont {Hashimoto}\ \emph {et~al.}(2014)\citenamefont
  {Hashimoto}, \citenamefont {Oka},\ and\ \citenamefont
  {Sonoda}}]{Hashimoto:2014dza}%
  \BibitemOpen
  \bibfield  {author} {\bibinfo {author} {\bibfnamefont {Koji}\ \bibnamefont
  {Hashimoto}}, \bibinfo {author} {\bibfnamefont {Takashi}\ \bibnamefont
  {Oka}}, \ and\ \bibinfo {author} {\bibfnamefont {Akihiko}\ \bibnamefont
  {Sonoda}},\ }\bibfield  {title} {\enquote {\bibinfo {title} {{Magnetic
  instability in AdS/CFT: Schwinger effect and Euler-Heisenberg Lagrangian of
  supersymmetric QCD}},}\ }\href {\doibase 10.1007/JHEP06(2014)085} {\bibfield
  {journal} {\bibinfo  {journal} {JHEP}\ }\textbf {\bibinfo {volume} {06}},\
  \bibinfo {pages} {085} (\bibinfo {year} {2014})},\ \Eprint
  {http://arxiv.org/abs/1403.6336} {arXiv:1403.6336 [hep-th]} \BibitemShut
  {NoStop}%
\bibitem [{\citenamefont {Karch}\ and\ \citenamefont
  {O'Bannon}(2007)}]{Karch:2007pd}%
  \BibitemOpen
  \bibfield  {author} {\bibinfo {author} {\bibfnamefont {Andreas}\ \bibnamefont
  {Karch}}\ and\ \bibinfo {author} {\bibfnamefont {Andy}\ \bibnamefont
  {O'Bannon}},\ }\bibfield  {title} {\enquote {\bibinfo {title} {{Metallic
  AdS/CFT}},}\ }\href {\doibase 10.1088/1126-6708/2007/09/024} {\bibfield
  {journal} {\bibinfo  {journal} {JHEP}\ }\textbf {\bibinfo {volume} {09}},\
  \bibinfo {pages} {024} (\bibinfo {year} {2007})},\ \Eprint
  {http://arxiv.org/abs/0705.3870} {arXiv:0705.3870 [hep-th]} \BibitemShut
  {NoStop}%
\bibitem [{\citenamefont {Chang}\ \emph {et~al.}(2014)\citenamefont {Chang},
  \citenamefont {Karch},\ and\ \citenamefont {Uhlemann}}]{Chang:2014oia}%
  \BibitemOpen
  \bibfield  {author} {\bibinfo {author} {\bibfnamefont {Han-Chih}\
  \bibnamefont {Chang}}, \bibinfo {author} {\bibfnamefont {Andreas}\
  \bibnamefont {Karch}}, \ and\ \bibinfo {author} {\bibfnamefont
  {Christoph~F.}\ \bibnamefont {Uhlemann}},\ }\bibfield  {title} {\enquote
  {\bibinfo {title} {{Flavored $\mathcal{N}=4$ SYM - a highly entangled quantum
  liquid}},}\ }\href {\doibase 10.1007/JHEP09(2014)110} {\bibfield  {journal}
  {\bibinfo  {journal} {JHEP}\ }\textbf {\bibinfo {volume} {09}},\ \bibinfo
  {pages} {110} (\bibinfo {year} {2014})},\ \Eprint
  {http://arxiv.org/abs/1406.2705} {arXiv:1406.2705 [hep-th]} \BibitemShut
  {NoStop}%
\bibitem [{\citenamefont {Freedman}\ \emph {et~al.}(1999)\citenamefont
  {Freedman}, \citenamefont {Gubser}, \citenamefont {Pilch},\ and\
  \citenamefont {Warner}}]{Freedman:1999gp}%
  \BibitemOpen
  \bibfield  {author} {\bibinfo {author} {\bibfnamefont {D.~Z.}\ \bibnamefont
  {Freedman}}, \bibinfo {author} {\bibfnamefont {S.~S.}\ \bibnamefont
  {Gubser}}, \bibinfo {author} {\bibfnamefont {K.}~\bibnamefont {Pilch}}, \
  and\ \bibinfo {author} {\bibfnamefont {N.~P.}\ \bibnamefont {Warner}},\
  }\bibfield  {title} {\enquote {\bibinfo {title} {{Renormalization group flows
  from holography supersymmetry and a c theorem}},}\ }\href@noop {} {\bibfield
  {journal} {\bibinfo  {journal} {Adv. Theor. Math. Phys.}\ }\textbf {\bibinfo
  {volume} {3}},\ \bibinfo {pages} {363--417} (\bibinfo {year} {1999})},\
  \Eprint {http://arxiv.org/abs/hep-th/9904017} {arXiv:hep-th/9904017 [hep-th]}
  \BibitemShut {NoStop}%
\bibitem [{\citenamefont {Kastor}\ \emph {et~al.}(2014)\citenamefont {Kastor},
  \citenamefont {Ray},\ and\ \citenamefont {Traschen}}]{Kastor:2014dra}%
  \BibitemOpen
  \bibfield  {author} {\bibinfo {author} {\bibfnamefont {David}\ \bibnamefont
  {Kastor}}, \bibinfo {author} {\bibfnamefont {Sourya}\ \bibnamefont {Ray}}, \
  and\ \bibinfo {author} {\bibfnamefont {Jennie}\ \bibnamefont {Traschen}},\
  }\bibfield  {title} {\enquote {\bibinfo {title} {{Chemical Potential in the
  First Law for Holographic Entanglement Entropy}},}\ }\href {\doibase
  10.1007/JHEP11(2014)120} {\bibfield  {journal} {\bibinfo  {journal} {JHEP}\
  }\textbf {\bibinfo {volume} {11}},\ \bibinfo {pages} {120} (\bibinfo {year}
  {2014})},\ \Eprint {http://arxiv.org/abs/1409.3521} {arXiv:1409.3521
  [hep-th]} \BibitemShut {NoStop}%
\bibitem [{\citenamefont {Das}\ \emph {et~al.}(2010)\citenamefont {Das},
  \citenamefont {Nishioka},\ and\ \citenamefont {Takayanagi}}]{Das:2010yw}%
  \BibitemOpen
  \bibfield  {author} {\bibinfo {author} {\bibfnamefont {Sumit~R.}\
  \bibnamefont {Das}}, \bibinfo {author} {\bibfnamefont {Tatsuma}\ \bibnamefont
  {Nishioka}}, \ and\ \bibinfo {author} {\bibfnamefont {Tadashi}\ \bibnamefont
  {Takayanagi}},\ }\bibfield  {title} {\enquote {\bibinfo {title} {{Probe
  Branes, Time-dependent Couplings and Thermalization in AdS/CFT}},}\ }\href
  {\doibase 10.1007/JHEP07(2010)071} {\bibfield  {journal} {\bibinfo  {journal}
  {JHEP}\ }\textbf {\bibinfo {volume} {07}},\ \bibinfo {pages} {071} (\bibinfo
  {year} {2010})},\ \Eprint {http://arxiv.org/abs/1005.3348} {arXiv:1005.3348
  [hep-th]} \BibitemShut {NoStop}%
\bibitem [{\citenamefont {Erdmenger}\ \emph {et~al.}(2007)\citenamefont
  {Erdmenger}, \citenamefont {Meyer},\ and\ \citenamefont
  {Shock}}]{Erdmenger:2007bn}%
  \BibitemOpen
  \bibfield  {author} {\bibinfo {author} {\bibfnamefont {Johanna}\ \bibnamefont
  {Erdmenger}}, \bibinfo {author} {\bibfnamefont {Rene}\ \bibnamefont {Meyer}},
  \ and\ \bibinfo {author} {\bibfnamefont {Jonathan~P.}\ \bibnamefont
  {Shock}},\ }\bibfield  {title} {\enquote {\bibinfo {title} {{AdS/CFT with
  Flavour in Electric and Magnetic Kalb-Ramond Fields}},}\ }\href {\doibase
  10.1088/1126-6708/2007/12/091} {\bibfield  {journal} {\bibinfo  {journal}
  {JHEP}\ }\textbf {\bibinfo {volume} {12}},\ \bibinfo {pages} {091} (\bibinfo
  {year} {2007})},\ \Eprint {http://arxiv.org/abs/0709.1551} {arXiv:0709.1551
  [hep-th]} \BibitemShut {NoStop}%
\bibitem [{\citenamefont {Albash}\ \emph {et~al.}(2008)\citenamefont {Albash},
  \citenamefont {Filev}, \citenamefont {Johnson},\ and\ \citenamefont
  {Kundu}}]{Albash:2007bq}%
  \BibitemOpen
  \bibfield  {author} {\bibinfo {author} {\bibfnamefont {Tameem}\ \bibnamefont
  {Albash}}, \bibinfo {author} {\bibfnamefont {Veselin~G.}\ \bibnamefont
  {Filev}}, \bibinfo {author} {\bibfnamefont {Clifford~V.}\ \bibnamefont
  {Johnson}}, \ and\ \bibinfo {author} {\bibfnamefont {Arnab}\ \bibnamefont
  {Kundu}},\ }\bibfield  {title} {\enquote {\bibinfo {title} {{Quarks in an
  External Electric Field in Finite Temperature Large N Gauge Theory}},}\
  }\href {\doibase 10.1088/1126-6708/2008/08/092} {\bibfield  {journal}
  {\bibinfo  {journal} {JHEP}\ }\textbf {\bibinfo {volume} {08}},\ \bibinfo
  {pages} {092} (\bibinfo {year} {2008})},\ \Eprint
  {http://arxiv.org/abs/0709.1554} {arXiv:0709.1554 [hep-th]} \BibitemShut
  {NoStop}%
\bibitem [{\citenamefont {Filev}\ and\ \citenamefont
  {Johnson}(2008)}]{Filev:2008xt}%
  \BibitemOpen
  \bibfield  {author} {\bibinfo {author} {\bibfnamefont {Veselin~G.}\
  \bibnamefont {Filev}}\ and\ \bibinfo {author} {\bibfnamefont {Clifford~V.}\
  \bibnamefont {Johnson}},\ }\bibfield  {title} {\enquote {\bibinfo {title}
  {{Universality in the Large N(c) Dynamics of Flavour: Thermal Vs. Quantum
  Induced Phase Transitions}},}\ }\href {\doibase
  10.1088/1126-6708/2008/10/058} {\bibfield  {journal} {\bibinfo  {journal}
  {JHEP}\ }\textbf {\bibinfo {volume} {10}},\ \bibinfo {pages} {058} (\bibinfo
  {year} {2008})},\ \Eprint {http://arxiv.org/abs/0805.1950} {arXiv:0805.1950
  [hep-th]} \BibitemShut {NoStop}%
\bibitem [{\citenamefont {Kim}\ \emph {et~al.}(2011)\citenamefont {Kim},
  \citenamefont {Shock},\ and\ \citenamefont {Tarrio}}]{Kim:2011qh}%
  \BibitemOpen
  \bibfield  {author} {\bibinfo {author} {\bibfnamefont {Keun-Young}\
  \bibnamefont {Kim}}, \bibinfo {author} {\bibfnamefont {Jonathan~P.}\
  \bibnamefont {Shock}}, \ and\ \bibinfo {author} {\bibfnamefont {Javier}\
  \bibnamefont {Tarrio}},\ }\bibfield  {title} {\enquote {\bibinfo {title}
  {{The open string membrane paradigm with external electromagnetic fields}},}\
  }\href {\doibase 10.1007/JHEP06(2011)017} {\bibfield  {journal} {\bibinfo
  {journal} {JHEP}\ }\textbf {\bibinfo {volume} {06}},\ \bibinfo {pages} {017}
  (\bibinfo {year} {2011})},\ \Eprint {http://arxiv.org/abs/1103.4581}
  {arXiv:1103.4581 [hep-th]} \BibitemShut {NoStop}%
\bibitem [{\citenamefont {Sonner}\ and\ \citenamefont
  {Green}(2012)}]{Sonner:2012if}%
  \BibitemOpen
  \bibfield  {author} {\bibinfo {author} {\bibfnamefont {Julian}\ \bibnamefont
  {Sonner}}\ and\ \bibinfo {author} {\bibfnamefont {Andrew~G.}\ \bibnamefont
  {Green}},\ }\bibfield  {title} {\enquote {\bibinfo {title} {{Hawking
  Radiation and Non-equilibrium Quantum Critical Current Noise}},}\ }\href
  {\doibase 10.1103/PhysRevLett.109.091601} {\bibfield  {journal} {\bibinfo
  {journal} {Phys. Rev. Lett.}\ }\textbf {\bibinfo {volume} {109}},\ \bibinfo
  {pages} {091601} (\bibinfo {year} {2012})},\ \Eprint
  {http://arxiv.org/abs/1203.4908} {arXiv:1203.4908 [cond-mat.str-el]}
  \BibitemShut {NoStop}%
\bibitem [{\citenamefont {Nakamura}\ and\ \citenamefont
  {Ooguri}(2013)}]{Nakamura:2013yqa}%
  \BibitemOpen
  \bibfield  {author} {\bibinfo {author} {\bibfnamefont {Shin}\ \bibnamefont
  {Nakamura}}\ and\ \bibinfo {author} {\bibfnamefont {Hirosi}\ \bibnamefont
  {Ooguri}},\ }\bibfield  {title} {\enquote {\bibinfo {title} {{Out of
  Equilibrium Temperature from Holography}},}\ }\href {\doibase
  10.1103/PhysRevD.88.126003} {\bibfield  {journal} {\bibinfo  {journal} {Phys.
  Rev.}\ }\textbf {\bibinfo {volume} {D88}},\ \bibinfo {pages} {126003}
  (\bibinfo {year} {2013})},\ \Eprint {http://arxiv.org/abs/1309.4089}
  {arXiv:1309.4089 [hep-th]} \BibitemShut {NoStop}%
\bibitem [{\citenamefont {Kundu}\ and\ \citenamefont
  {Kundu}(2015)}]{Kundu:2013eba}%
  \BibitemOpen
  \bibfield  {author} {\bibinfo {author} {\bibfnamefont {Arnab}\ \bibnamefont
  {Kundu}}\ and\ \bibinfo {author} {\bibfnamefont {Sandipan}\ \bibnamefont
  {Kundu}},\ }\bibfield  {title} {\enquote {\bibinfo {title} {{Steady-state
  Physics, Effective Temperature Dynamics in Holography}},}\ }\href {\doibase
  10.1103/PhysRevD.91.046004} {\bibfield  {journal} {\bibinfo  {journal} {Phys.
  Rev.}\ }\textbf {\bibinfo {volume} {D91}},\ \bibinfo {pages} {046004}
  (\bibinfo {year} {2015})},\ \Eprint {http://arxiv.org/abs/1307.6607}
  {arXiv:1307.6607 [hep-th]} \BibitemShut {NoStop}%
\bibitem [{\citenamefont {Kundu}(2015)}]{Kundu:2015qda}%
  \BibitemOpen
  \bibfield  {author} {\bibinfo {author} {\bibfnamefont {Arnab}\ \bibnamefont
  {Kundu}},\ }\bibfield  {title} {\enquote {\bibinfo {title} {{Effective
  Temperature in Steady-state Dynamics from Holography}},}\ }\href {\doibase
  10.1007/JHEP09(2015)042} {\bibfield  {journal} {\bibinfo  {journal} {JHEP}\
  }\textbf {\bibinfo {volume} {09}},\ \bibinfo {pages} {042} (\bibinfo {year}
  {2015})},\ \Eprint {http://arxiv.org/abs/1507.00818} {arXiv:1507.00818
  [hep-th]} \BibitemShut {NoStop}%
\bibitem [{\citenamefont {Banerjee}\ \emph
  {et~al.}(2016{\natexlab{a}})\citenamefont {Banerjee}, \citenamefont {Kundu},\
  and\ \citenamefont {Kundu}}]{Banerjee:2015cvy}%
  \BibitemOpen
  \bibfield  {author} {\bibinfo {author} {\bibfnamefont {Avik}\ \bibnamefont
  {Banerjee}}, \bibinfo {author} {\bibfnamefont {Arnab}\ \bibnamefont {Kundu}},
  \ and\ \bibinfo {author} {\bibfnamefont {Sandipan}\ \bibnamefont {Kundu}},\
  }\bibfield  {title} {\enquote {\bibinfo {title} {{Flavour Fields in Steady
  State: Stress Tensor and Free Energy}},}\ }\href {\doibase
  10.1007/JHEP02(2016)102} {\bibfield  {journal} {\bibinfo  {journal} {JHEP}\
  }\textbf {\bibinfo {volume} {02}},\ \bibinfo {pages} {102} (\bibinfo {year}
  {2016}{\natexlab{a}})},\ \Eprint {http://arxiv.org/abs/1512.05472}
  {arXiv:1512.05472 [hep-th]} \BibitemShut {NoStop}%
\bibitem [{\citenamefont {Banerjee}\ \emph
  {et~al.}(2016{\natexlab{b}})\citenamefont {Banerjee}, \citenamefont {Kundu},\
  and\ \citenamefont {Kundu}}]{Banerjee:2016qeu}%
  \BibitemOpen
  \bibfield  {author} {\bibinfo {author} {\bibfnamefont {Avik}\ \bibnamefont
  {Banerjee}}, \bibinfo {author} {\bibfnamefont {Arnab}\ \bibnamefont {Kundu}},
  \ and\ \bibinfo {author} {\bibfnamefont {Sandipan}\ \bibnamefont {Kundu}},\
  }\bibfield  {title} {\enquote {\bibinfo {title} {{Emergent Horizons and
  Causal Structures in Holography}},}\ }\href {\doibase
  10.1007/JHEP09(2016)166} {\bibfield  {journal} {\bibinfo  {journal} {JHEP}\
  }\textbf {\bibinfo {volume} {09}},\ \bibinfo {pages} {166} (\bibinfo {year}
  {2016}{\natexlab{b}})},\ \Eprint {http://arxiv.org/abs/1605.07368}
  {arXiv:1605.07368 [hep-th]} \BibitemShut {NoStop}%
\bibitem [{\citenamefont {Sonner}(2013)}]{Sonner:2013mba}%
  \BibitemOpen
  \bibfield  {author} {\bibinfo {author} {\bibfnamefont {Julian}\ \bibnamefont
  {Sonner}},\ }\bibfield  {title} {\enquote {\bibinfo {title} {{Holographic
  Schwinger Effect and the Geometry of Entanglement}},}\ }\href {\doibase
  10.1103/PhysRevLett.111.211603} {\bibfield  {journal} {\bibinfo  {journal}
  {Phys. Rev. Lett.}\ }\textbf {\bibinfo {volume} {111}},\ \bibinfo {pages}
  {211603} (\bibinfo {year} {2013})},\ \Eprint {http://arxiv.org/abs/1307.6850}
  {arXiv:1307.6850 [hep-th]} \BibitemShut {NoStop}%
\bibitem [{\citenamefont {Aharony}\ and\ \citenamefont
  {Razamat}(2002)}]{Aharony:2002tp}%
  \BibitemOpen
  \bibfield  {author} {\bibinfo {author} {\bibfnamefont {Ofer}\ \bibnamefont
  {Aharony}}\ and\ \bibinfo {author} {\bibfnamefont {Shlomo~S.}\ \bibnamefont
  {Razamat}},\ }\bibfield  {title} {\enquote {\bibinfo {title} {{Exactly
  marginal deformations of N=4 SYM and of its supersymmetric orbifold
  descendants}},}\ }\href {\doibase 10.1088/1126-6708/2002/05/029} {\bibfield
  {journal} {\bibinfo  {journal} {JHEP}\ }\textbf {\bibinfo {volume} {05}},\
  \bibinfo {pages} {029} (\bibinfo {year} {2002})},\ \Eprint
  {http://arxiv.org/abs/hep-th/0204045} {arXiv:hep-th/0204045 [hep-th]}
  \BibitemShut {NoStop}%
\bibitem [{\citenamefont {Aharony}\ \emph {et~al.}(2002)\citenamefont
  {Aharony}, \citenamefont {Kol},\ and\ \citenamefont
  {Yankielowicz}}]{Aharony:2002hx}%
  \BibitemOpen
  \bibfield  {author} {\bibinfo {author} {\bibfnamefont {Ofer}\ \bibnamefont
  {Aharony}}, \bibinfo {author} {\bibfnamefont {Barak}\ \bibnamefont {Kol}}, \
  and\ \bibinfo {author} {\bibfnamefont {Shimon}\ \bibnamefont
  {Yankielowicz}},\ }\bibfield  {title} {\enquote {\bibinfo {title} {{On
  exactly marginal deformations of N=4 SYM and type IIB supergravity on AdS(5)
  x S**5}},}\ }\href {\doibase 10.1088/1126-6708/2002/06/039} {\bibfield
  {journal} {\bibinfo  {journal} {JHEP}\ }\textbf {\bibinfo {volume} {06}},\
  \bibinfo {pages} {039} (\bibinfo {year} {2002})},\ \Eprint
  {http://arxiv.org/abs/hep-th/0205090} {arXiv:hep-th/0205090 [hep-th]}
  \BibitemShut {NoStop}%
\bibitem [{\citenamefont {de~Haro}\ \emph
  {et~al.}(2001{\natexlab{b}})\citenamefont {de~Haro}, \citenamefont
  {Solodukhin},\ and\ \citenamefont {Skenderis}}]{deHaro:2000xn}%
  \BibitemOpen
  \bibfield  {author} {\bibinfo {author} {\bibfnamefont {Sebastian}\
  \bibnamefont {de~Haro}}, \bibinfo {author} {\bibfnamefont {Sergey~N.}\
  \bibnamefont {Solodukhin}}, \ and\ \bibinfo {author} {\bibfnamefont {Kostas}\
  \bibnamefont {Skenderis}},\ }\bibfield  {title} {\enquote {\bibinfo {title}
  {{Holographic Reconstruction of Spacetime and Renormalization in the AdS/CFT
  Correspondence}},}\ }\href {\doibase 10.1007/s002200100381} {\bibfield
  {journal} {\bibinfo  {journal} {Commun. Math. Phys.}\ }\textbf {\bibinfo
  {volume} {217}},\ \bibinfo {pages} {595--622} (\bibinfo {year}
  {2001}{\natexlab{b}})},\ \Eprint {http://arxiv.org/abs/hep-th/0002230}
  {arXiv:hep-th/0002230} \BibitemShut {NoStop}%
\bibitem [{\citenamefont {Reza Mohammadi~Mozaffar}\ \emph
  {et~al.}(2016)\citenamefont {Reza Mohammadi~Mozaffar}, \citenamefont
  {Mollabashi},\ and\ \citenamefont {Omidi}}]{Mozaffara:2016iwm}%
  \BibitemOpen
  \bibfield  {author} {\bibinfo {author} {\bibfnamefont {M.}~\bibnamefont {Reza
  Mohammadi~Mozaffar}}, \bibinfo {author} {\bibfnamefont {Ali}\ \bibnamefont
  {Mollabashi}}, \ and\ \bibinfo {author} {\bibfnamefont {Farzad}\ \bibnamefont
  {Omidi}},\ }\bibfield  {title} {\enquote {\bibinfo {title} {{Non-local Probes
  in Holographic Theories with Momentum Relaxation}},}\ }\href {\doibase
  10.1007/JHEP10(2016)135} {\bibfield  {journal} {\bibinfo  {journal} {JHEP}\
  }\textbf {\bibinfo {volume} {10}},\ \bibinfo {pages} {135} (\bibinfo {year}
  {2016})},\ \Eprint {http://arxiv.org/abs/1608.08781} {arXiv:1608.08781
  [hep-th]} \BibitemShut {NoStop}%
\end{thebibliography}%

\end{document}